\documentclass{aa}  
\usepackage{graphicx}
\usepackage{txfonts}
\newcommand{\mincir}{\raise -2.truept\hbox{\rlap{\hbox{$\sim$}}\raise5.truept
\hbox{$<$}\ }}
\newcommand{\magcir}{\raise -2.truept\hbox{\rlap{\hbox{$\sim$}}\raise5.truept
\hbox{$>$}\ }}
\newcommand{\siml}{\raise -2.truept\hbox{\rlap{\hbox{$\sim$}}\raise5.truept
\hbox{$<$}\ }}
\newcommand{\simg}{\raise -2.truept\hbox{\rlap{\hbox{$\sim$}}\raise5.truept
\hbox{$>$}\ }}
\newcommand{\be}{\begin{equation}}
\newcommand{\ee}{\end{equation}}
\newcommand{\ba}{\begin{eqnarray}}
\newcommand{\ea}{\end{eqnarray}}
\newcommand {\kpc} {$\mathrm{h_{70}^{-1}}$ kpc $\;$}

\newcommand {\h} {$h_{70}^{-1}$ Mpc$\;$}
\newcommand {\hh} {$h_{70}^{-1}$ Mpc}
\newcommand {\hhh} {\;h_{70}^{-1} \mathrm{Mpc}}
\newcommand {\ks} {km~s$^{-1} \;$}
\newcommand {\kss} {km~s$^{-1}$}

\newcommand {\mqua} {$\times 10^{14}\;h_{70}^{-1}\;M_{\odot} \;$}
\newcommand {\mquaa} {$\times 10^{14}\;h_{70}^{-1}\;M_{\odot}$}
\newcommand {\mqui} {$\times 10^{15}\;h_{70}^{-1}\;M_{\odot} \;$}
\newcommand {\mquii} {$\times 10^{15}\;h_{70}^{-1}\;M_{\odot}$}

\newcommand{\degree}{\ensuremath{\mathrm{^\circ}}}
\newcommand{\arcm}{\ensuremath{\mathrm{^\prime}\;}}
\newcommand{\arcs}{\ensuremath{\arcmm\hskip -0.1em\arcmm \;}}
\newcommand{\arcmm}{\ensuremath{\mathrm{^\prime}}}
\newcommand{\arcss}{\ensuremath{\arcmm\hskip -0.1em\arcmm}}

\newcommand{\dotsec}{\,\rlap{\hbox{$\mathrm{^s}$}}{\hbox{$.$}}\,}

\begin{document}
   \title{Internal dynamics of Abell 1240: a galaxy cluster with
symmetric double radio relics}

   \author{
R. Barrena\inst{1}
          \and
M. Girardi\inst{2,3}
          \and
W. Boschin\inst{4,2}
          \and
M. Das\'i\inst{5,1}
}

   \offprints{R. Barrena, \email{rbarrena@iac.es}}

   \institute{Instituto de Astrof\'{\i}sica de Canarias,
	      C/V\'{\i}a L\'actea s/n, E-38205 La Laguna (Tenerife),
	      Canary Islands, Spain\\
\and 
              Dipartimento di Astronomia of the Universit\`a degli
	      Studi di Trieste, via Tiepolo 11, I-34143 Trieste,
	      Italy\\ 
\and          
              INAF - Osservatorio Astronomico di Trieste, via Tiepolo 11,
              I-34143 Trieste, Italy\\
\and
	      Fundaci\'on Galileo Galilei - INAF, Rambla Jos\'e Ana 
              Fern\'andez Perez 7, E-38712 Bre\~na Baja (La Palma), 
              Canary Islands, Spain\\
\and
              Max-Planck-Institut f\"ur Sonnensystemforschung, 
	      Max-Planck-Str. 2, G-37191 Katlenburg-Lindau, Germany\\ }

\date{Received  / Accepted }

\abstract {The mechanisms giving rise to diffuse radio emission in
  galaxy clusters, and in particular their connection with cluster
  mergers, are still debated.}{We aim to obtain new insights into the
  internal dynamics of the cluster Abell 1240, showing the presence of
  two roughly symmetric radio relics, separated by $\sim 2$ \hh.}{Our
  analysis is mainly based on redshift data for 145 galaxies
  mostly acquired at the Telescopio Nazionale Galileo and on new
  photometric data acquired at the Isaac Newton Telescope. We also use
  X--ray data from the Chandra archive and photometric data from the
  Sloan Digital Sky Survey (Data Release 7). We combine galaxy
  velocities and positions to select 89 cluster galaxies and analyze
  the internal dynamics of the Abell 1237 + Abell 1240 cluster
  complex, being Abell 1237 a close companion of Abell 1240 towards
  the southern direction.} {We estimate similar redshifts for Abell
  1237 and Abell 1240, $\left<z\right>=0.1935$ and
  $\left<z\right>=0.1948$, respectively. For Abell 1237 we estimate a
  line--of--sight (LOS) velocity dispersion $\sigma_{\rm V}\sim 740$
  \ks and a mass $M\sim 6$ \mquaa. For Abell 1240 we estimate a LOS
  $\sigma_{\rm V}\sim 870$ \ks and a mass range $M\sim 0.9-1.9$
  \mquii, which takes into account its complex dynamics. Abell 1240 is
  shown to have a bimodal structure with two galaxy clumps roughly
  defining the N--S direction, the same one defined by the elongation
  of its X--ray surface brightness and by the axis of symmetry of the
  relics. The two brightest galaxies of Abell 1240, associated to the
  northern and southern clumps, are separated by a LOS rest--frame
  velocity difference $V_{\rm rf}\sim 400$ \ks and by a projected
  distance $D\sim 1.2$ \hh.  The two--body model agrees with the
  hypothesis that we are looking at a cluster merger occurred largely
  in the plane of the sky, with the two galaxy clumps separated by a
  rest--frame velocity difference $V_{\rm rf}\sim 2000$ \ks at a time
  of 0.3 Gyrs after the crossing core, while Abell 1237 is still
  infalling onto Abell 1240. Chandra archive data confirm the complex
  structure of Abell 1240 and allow us to estimate a global X--ray
  temperature $T_{\rm X}=\,$6.0\,$\pm\,0.5$ keV.}{In agreement with
  the findings from radio data, our results for Abell 1240 strongly
  support the ``outgoing merger shocks'' model to explain the presence
  of the relics.}


   \titlerunning{Internal dynamics of Abell 1240}
   \maketitle
%

\section{Introduction}
\label{intr}

Merging processes constitute an essential ingredient of the evolution
of galaxy clusters (see Feretti et al. \cite{fer02b} for a review). An
interesting aspect of these phenomena is the possible connection of
cluster mergers with the presence of extended, diffuse radio sources:
halos and relics. The synchrotron radio emission of these sources
demonstrates the existence of large--scale cluster magnetic fields and
of widespread relativistic particles. Cluster mergers have been
suggested to provide the large amount of energy necessary for electron
reacceleration up to relativistic energies and for magnetic field
amplification (Feretti \cite{fer99}; Feretti \cite{fer02a}; Sarazin
\cite{sar02}). Radio relics (``radio gischts'' as referred by Kempner
et al. \cite{kem03}), which are polarized and elongated radio sources
located in the cluster peripheral regions, seem to be directly
associated with merger shocks (e.g., Ensslin et al. \cite{ens98};
Roettiger et al. \cite{roe99}; Ensslin \& Gopal--Krishna \cite{ens01};
Hoeft et al. \cite{hoe04}).  Radio halos, unpolarized sources which
permeate the cluster volume similarly to the X--ray emitting gas, are
more likely to be associated with the turbulence following a cluster
merger (Cassano \& Brunetti \cite{cas05}). However, the precise
radio halos/relics formation scenario is still debated since the
diffuse radio sources are quite uncommon and only recently one can
study these phenomena on the basis of a sufficient statistics (few
dozen clusters up to $z\sim 0.3$, e.g., Giovannini et
al. \cite{gio99}; see also Giovannini \& Feretti \cite{gio02}; Feretti
\cite{fer05a}) and attempt a classification (e.g., Kempner et
al. \cite{kem03}; Ferrari et al. \cite{ferr08}).

There is growing evidence of the connection between diffuse radio
emission and cluster merging, since up to now diffuse radio 
sources have been detected only in merging systems. In most of the 
cases the cluster dynamical state has been derived from X--ray 
observations (see Buote \cite{buo02}; Feretti \cite{fer06} and \cite{fer08} 
and refs. therein). Optical data are a powerful way to investigate the
presence and the dynamics of cluster mergers (e.g., Girardi \& Biviano
\cite{gir02}), too. The spatial and kinematical analysis of member
galaxies allow us to detect and measure the amount of substructure, to
identify and analyze possible pre--merging clumps or merger remnants.
This optical information is really complementary to X--ray information
since galaxies and intra--cluster medium react on different time
scales during a merger (see, e.g., numerical simulations by Roettiger
et al. \cite{roe97}). In this context we are conducting an intensive
observational and data analysis program to study the internal dynamics
of clusters with diffuse radio emission by using member galaxies 
(Girardi et al. \cite{gir07} and refs. therein \footnote{please visit 
the web site of the DARC (Dynamical Analysis of Radio Clusters) project:
http://adlibitum.oat.ts.astro.it/girardi/darc.}).

During our observational program we have conducted an intensive study
of the cluster \object{Abell 1240} (hereafter A1240).

A1240 is a very rich, X--ray luminous, Abell cluster: Abell richness
class $=2$ (Abell et al. \cite{abe89}); $L_\mathrm{X}$(0.1--2.0
keV)=8.3$\times 10^{43} \ h_{70}^{-2}$ erg\ s$^{-1}$ recovered from
ROSAT data (David et al. \cite{dav99}, correcting for our cluster
redshift, see below).  Optically, the cluster center is not dominated
by any single galaxy -- it is classified as Bautz--Morgan class III
(Abell et al. \cite{abe89}).

\begin{figure*}
\centering 
\includegraphics[width=18cm]{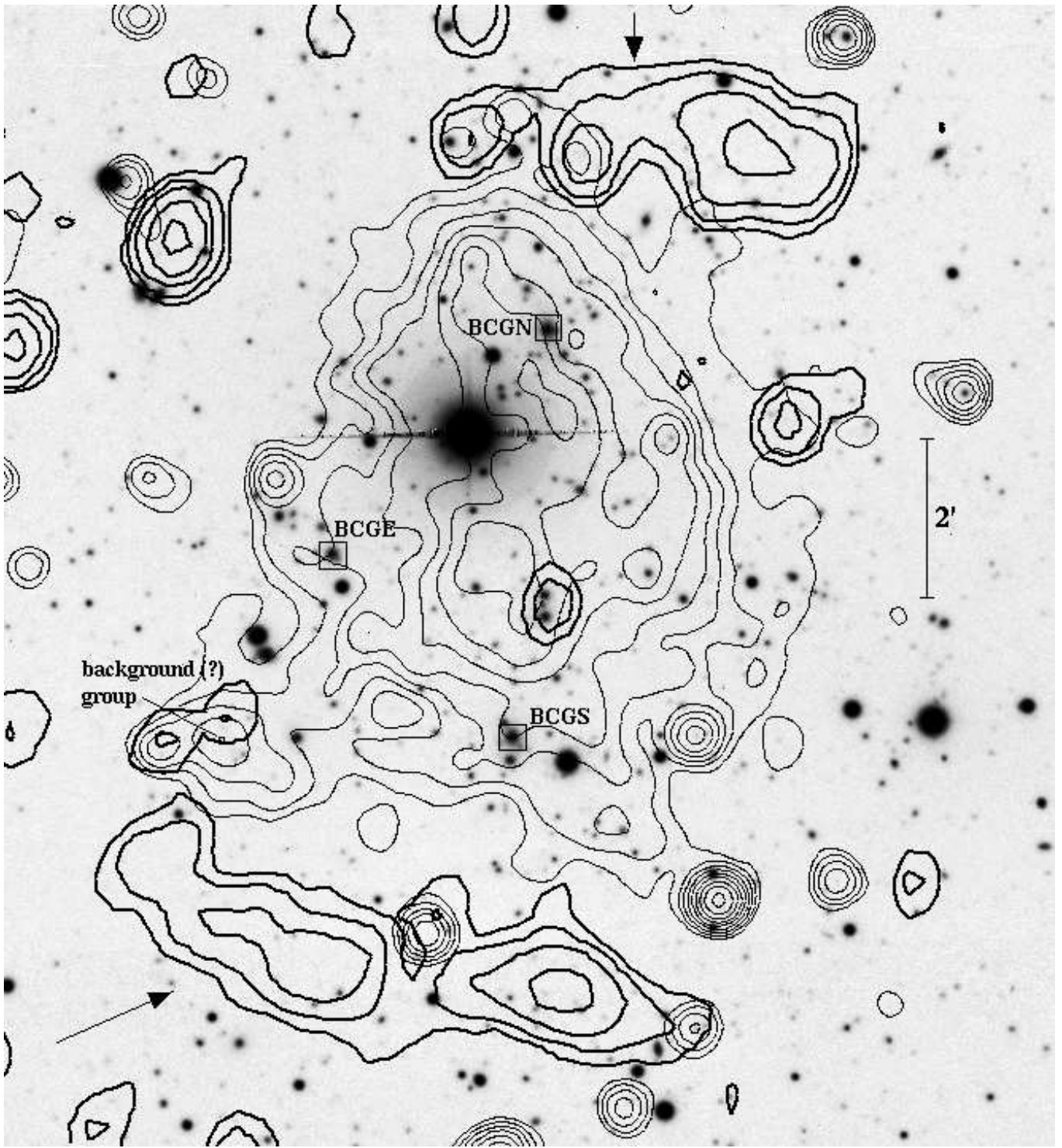}
\caption{INT $R$--band image of the cluster A1240 (North at the top
and East to the left) with, superimposed, the contour levels of the
Chandra archival image ID~4961 (thin contours; photons in the energy
range 0.5--2 keV) and the contour levels of a VLA radio image at 1.4
GHz (thick contours; Bonafede et al \cite{bon09}). Arrows show the
positions of the two radio relics. Boxes highlight the brightest
galaxies of A1240: BCGN, BCGS and BCGE (see text).}
\label{figimage}
\end{figure*}

Kempner \& Sarazin (\cite{kem01}) revealed the presence of two roughly
symmetric radio relics from the Westerbork Northern Sky Survey. They
appear to either side of the cluster center, north and south, at
distances of $\sim 6$\arcm and 7\arcmm. Kempner \& Sarazin also
noticed that A1240 shows an elongated X--ray morphology (recovered
from ROSAT observations) consistent with a slightly asymmetric merger
with the apparent axis roughly aligned with the axis of symmetry of
the relics (see also Bonafede et al. \cite{bon09}). The presence of
double relics was confirmed by recent, deep VLA observations (Bonafede
et al. \cite{bon09}; see Fig.~\ref{figimage}).  Very few other
clusters with double relics have been observed: Abell 3667 
(R\"ottgering et al. \cite{rot97}), Abell 3376 (Bagchi et
al. \cite{bag06}), Abell 2345 (Giovannini et al. \cite{gio99};
Bonafede et al. \cite{bon09}) and RXCJ 1314.4--2515 (Feretti et
al. \cite{fer05b}; Venturi et al. \cite{ven07}). The relics of Abell
3667 were explained with the ``outgoing merger shocks'' model
(Roettiger et al. \cite{roe99}). Observations of Abell 3376 agree with
both the ``outgoing merger shocks'' and the ``accretion shocks''
models (Bagchi et al. \cite{bag06}). In the case of Abell 2345 the
observations are difficult to reconcile with theoretical scenarios
(Bonafede et al. \cite{bon09}). Instead, more data are needed for RXCJ
1314.4--2515 (Feretti et al. \cite{fer05b} and Venturi et
al. \cite{ven07}). As for A1240, the detailed analysis of its radio
properties is in agreement with the ``outgoing merger shocks''
(Bonafede et al. \cite{bon09}), but the main global properties are
unknown and the internal cluster dynamics was never studied.

Indeed, few spectroscopic data have been reported in the field of
A1240 (see NED) and the value usually quoted in the literature for the
cluster redshift ($z=0.159$; see, e.g., David et al. \cite{dav99}) is
given by Ebeling et al. (\cite{ebe96}), on the basis of the
10th--ranked cluster galaxy. The real cluster redshift, as estimated in
this paper, is rather $z=0.195$.  Even poorer information is known for
\object{Abell 1237} (hereafter A1237), a close southern companion of
A1240, having richness class $=1$ and Bautz--Morgan class III (Abell
et al. \cite{abe89}).

Recently, we performed spectroscopic and photometric observations of
the A1237+A1240 complex with the Telescopio Nazionale Galileo (TNG)
and the Isaac Newton Telescope (INT), respectively. Our present
analysis is mainly based on our new optical data and X--ray Chandra
archival data. We also use the few public redshifts and photometric
data from the Sloan Digital Sky Survey (SDSS, Data Release 7). This
paper is organized as follows. We present our new optical data and the
cluster catalog in Sect.~2. We present our results about the cluster
structure based on optical and X--ray data in Sects.~3 and 4,
respectively. Finally, we briefly discuss our results and give our
conclusions in Sects.~5 and ~6.

Unless otherwise stated, we give errors at the 68\% confidence level
(hereafter c.l.). Results with a c.l. below $90\%$ are
considered very poorly/no significant. The values of these c.ls. are
generally not explicitly listed throughout the paper.

Throughout this paper, we use $H_0=70$ km s$^{-1}$
Mpc$^{-1}$ in a flat cosmology with $\Omega_0=0.3$ and
$\Omega_{\Lambda}=0.7$. In the adopted cosmology, 1\arcm corresponds
to $\sim 194$ \kpc at the cluster redshift.

\begin{figure*}
\centering 
\includegraphics[width=18cm]{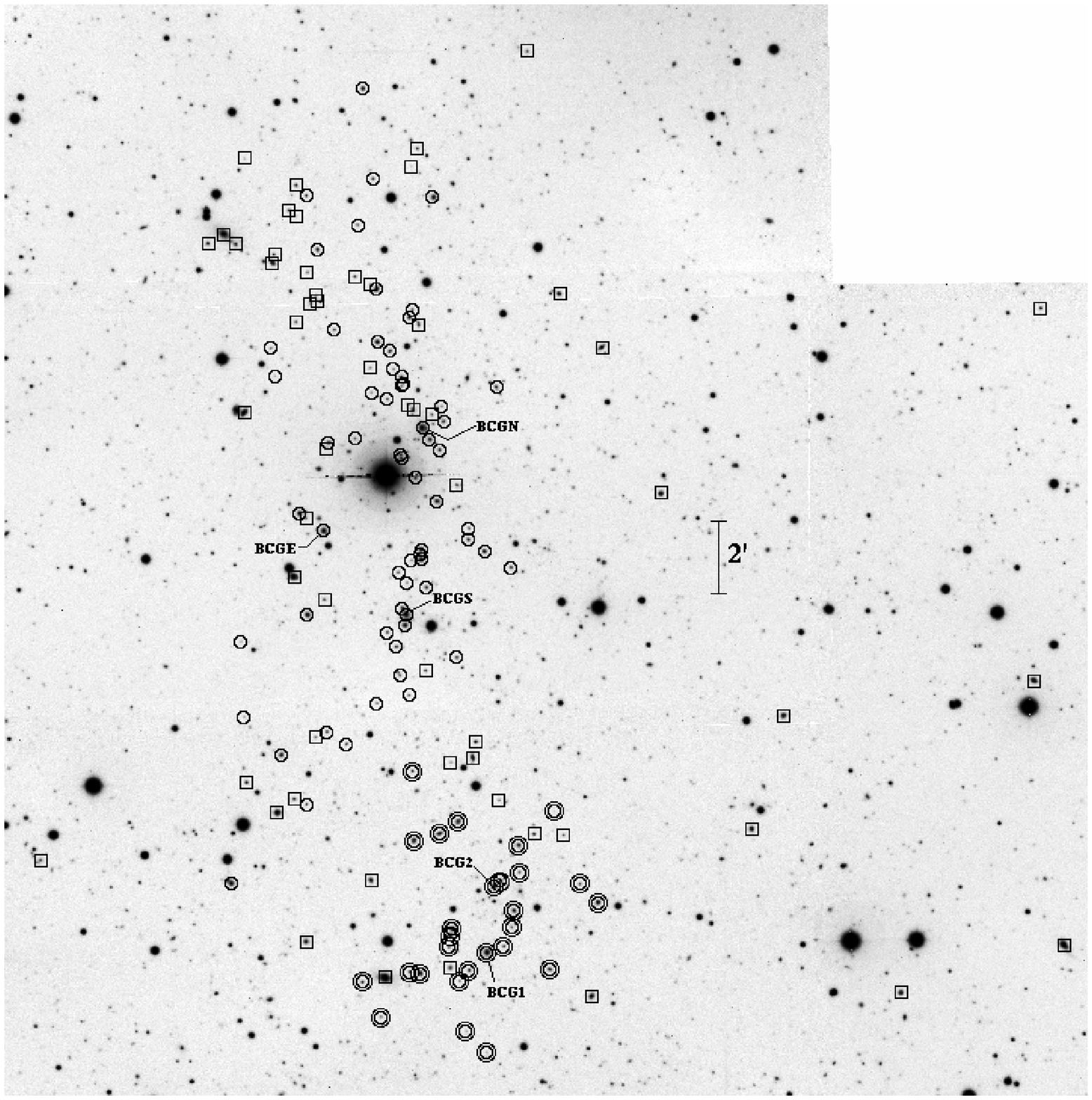}
\caption{INT $R$--band image of the A1237+A1240 complex (North at the
  top and East to the left). Circles and annuli indicate cluster
  members of A1240 and A1237, respectively (see
  Table~\ref{catalogA1240}). Boxes indicate non--member galaxies.}
\label{figottico}
\end{figure*}

\section{New data and galaxy catalog}
\label{newd}

Multi--object spectroscopic observations of A1240 were carried out at
the TNG telescope in December 2006 and December 2007. We used
DOLORES/MOS with the LR--B Grism 1, yielding a dispersion of 187
\AA/mm. In December 2006 we used the old Loral CCD, with a pixel size
of 15 $\mu$m, while in December 2007 we used the new E2V CCD, with a
pixel size of 13.5 $\mu$m. Both the CCDs are matrices of
$2048\times2048$ pixels. In total we observed four MOS masks (2 in
2006 and 2 in 2007) for a total of 142 slits.  We acquired three
exposures of 1800 s for each mask. Wavelength calibration was
performed using Helium--Argon lamps. Reduction of spectroscopic data
was carried out with the IRAF
\footnote{IRAF is distributed by the National Optical Astronomy
Observatories, which are operated by the Association of Universities
for Research in Astronomy, Inc., under cooperative agreement with the
National Science Foundation.} package.

Radial velocities were determined using the cross--correlation
technique (Tonry \& Davis \cite{ton79}) implemented in the RVSAO
package (developed at the Smithsonian Astrophysical Observatory
Telescope Data Center).  Each spectrum was correlated against six
templates for a variety of galaxy spectral types: E, S0, Sa, Sb, Sc,
Ir (Kennicutt \cite{ken92}). The template producing the highest value
of $\cal R$, i.e., the parameter given by RVSAO and related to the
signal--to--noise ratio of the correlation peak, was chosen. Moreover,
all spectra and their best correlation functions were examined
visually to verify the redshift determination. In 4 cases (IDs 55, 61,
69, 129, 137; see Table~\ref{catalogA1240}) we took the EMSAO redshift
as a reliable estimate of the redshift. Our spectroscopic survey in
the field of A1240 consists of spectra for 118 galaxies with a median
nominal error on $cz$ of 50 \kss.  The nominal errors as given by the
cross--correlation are known to be smaller than the true errors (e.g.,
Malumuth et al. \cite{mal92}; Bardelli et al. \cite{bar94}; Ellingson
\& Yee \cite{ell94}; Quintana et al. \cite{qui00}; Boschin et
al. \cite{bos04}). Double redshift determinations for four galaxies
allowed us to estimate real intrinsic errors. We compared the first
and second determinations computing the mean and the rms of the
variable $(z_1-z_2)/\sqrt{err_1^2+err_2^2}$.  We obtained
$mean=0.3\pm1.4$ and $rms=2.8$, to be compared with the expected
values of 0 and 1. The resulting mean shows that the two sets of
measurements are consistent with having the same velocity
zero--point. According to the $\chi^2$--test the high value of the rms
suggests that the errors are underestimated. Only when nominal errors
are multiplied by a $\sim 2$ factor the rms is in acceptable agreement
with the value of 1. Therefore, hereafter we assume that true errors
are larger than nominal cross--correlation errors by a factor 2. For
the four galaxies with two redshift estimates we used the weighted
mean of the two measurements and the corresponding errors.

We also use 32 public redshift data as taken from NED within a box of
45x45\arcm from the cluster center. They come from the SDSS (Data
Release 7). Before to proceed with the merging between our and
published catalogs we payed particular attention to their
compatibility. Five galaxies are in common between SDSS and TNG
data. For them we computed the mean and the rms of the variable
$(z_{\rm our}-z_{\rm lit})/\sqrt{err_{\rm our}^2+err_{\rm lit}^2}$. We
obtained $mean=-0.1\pm0.6$ and $rms=1.5$, to be compared with the
expected values of 0 and 1. The resulting mean shows that the two sets
of measurements are consistent with having the same velocity
zero--point, and the value of rms is compatible with a value of 1
according to the $\chi^2$--test. Thus we added the redshifts coming
from the literature. For the five galaxies in common we used the
weighted mean of the two redshift determinations and the corresponding
errors. We obtained a final catalog of 145 galaxies with measured
radial velocities.

As far as photometry is concerned, our observations were carried out
with the Wide Field Camera (WFC), mounted at the prime focus of the
2.5m INT telescope. We observed the A1237+A1240 complex with $B_{\rm
H}$ and $R_{\rm H}$ in photometric conditions. The $R$ image was
obtained in December 19th 2004 with a seeing condition of 3.0\arcss. We
got the $B$ image in May 14th 2006 with a seeing of about 1.1\arcss.

The WFC consists of a four--CCD mosaic covering a 33\arcmm$\times$33\arcm 
field of view, with only a 20\% marginally vignetted area. We took nine 
exposures of 720 s in $B_{\rm H}$ and 300 s in $R_{\rm H}$ Harris filters 
(a total of 6480 s and 2700 s in each band) developing a dithering 
pattern of nine positions. This observing mode allowed us to build a 
``supersky'' frame that was used to correct our images for fringing 
patterns (Gullixson \cite{gul92}). In addition, the dithering helped us 
to clean cosmic rays and avoid gaps between the CCDs in the final images. 
The complete reduction process (including flat fielding, bias subtraction 
and bad--column elimination) yielded a final coadded image where the 
variation of the sky was lower than 0.4\% in the whole frame. Another 
effect associated with the wide field frames is the distortion of the 
field. In order to match the photometry of several filters, a good 
astrometric solution is needed to take into account these distortions. 
Using the $imcoords$ IRAF tasks and taking as a reference the USNO B1.0 
catalog, we were able to find an accurate astrometric solution 
(rms $\sim$ 0.5\arcss) across the full frame. The photometric calibration 
was performed using Landolt standard fields obtained during the 
observation.

We finally identified galaxies in our $B_{\rm H}$ and $R_{\rm H}$
images and measured their magnitudes with the SExtractor package
(Bertin \& Arnouts \cite{ber96}) and AUTOMAG procedure. In a few cases
(e.g.\ close companion galaxies, galaxies close to defects of the CCD)
the standard SExtractor photometric procedure failed. In these cases
we computed magnitudes by hand. This method consists in assuming a
galaxy profile of a typical elliptical galaxy and scaling it to the
maximum observed value. The integration of this profile gives us an
estimate of the magnitude. This method is similar to PSF photometry,
but assumes a galaxy profile, more appropriate in this case.

We transformed all magnitudes into the Johnson--Cousins system
(Johnson \& Morgan \cite{joh53}; Cousins \cite{cou76}). We used
$B=B\rm_H+0.13$ and $R=R\rm_H$ as derived from the Harris filter
characterization
(http://www.ast.cam.ac.uk/$\sim$wfcsur/technical/photom/colours/) and
assuming a $B-V\sim 1.0$ for E--type galaxies (Poggianti \cite{pog97}).
As a final step, we estimated and corrected the galactic extinction
$A_B \sim0.10$, $A_R \sim0.06$ from Burstein \& Heiles's
(\cite{bur82}) reddening maps.

We estimated that our photometric sample is complete down to $R=20.5$
(22.0) and $B=22.0$ (23.0) for $S/N=5$ (3) within the observed field.

We assigned magnitudes to all galaxies of our spectroscopic catalog.
We measured redshifts for galaxies down to magnitude $R\sim$ 20, but a
high level of completeness is reached only for galaxies with magnitude
$R<$ 19 ($\sim$45\% completeness).

Table~\ref{catalogA1240} lists the velocity catalog (see also
Fig.~\ref{figottico}): identification number of each galaxy, ID
(Col.~1); right ascension and declination, $\alpha$ and $\delta$
(J2000, Col.~2); B and R magnitudes (Cols.~3 and 4); heliocentric
radial velocities, ${\rm v}=cz_{\sun}$ (Col.~5) with errors, $\Delta
{\rm v}$ (Col.~6); redshift source (Col.~7; T:TNG, S:SDSS); member
assignment (Col.~8; 1:A1240, 2:A1237, 0:background/foreground). 

\newcommand{\tng}{\mathrm{T}}
\newcommand{\sds}{\mathrm{S}}
\newcommand{\tns}{\mathrm{T+S}}

\begin{table}[!ht]
        \caption[]{Velocity catalog of 145 spectroscopically measured
galaxies in the field of the A1237+A1240 complex. In Col.~1, IDs 75, 56, 111, 29 and 27 (in boldface) highlight BCGS, BCGN, BCGE, BCG1 and BCG2, respectively (see text). Asterisks in Col.~8 highlight emission line galaxies.}
         \label{catalogA1240}
              $$ 
           \begin{array}{r c c r r r c r}
            \hline
            \noalign{\smallskip}
            \hline
            \noalign{\smallskip}

\mathrm{ID} & \mathrm{\alpha},\mathrm{\delta}\,(\mathrm{J}2000)  & B & R &\mathrm{v}\,\,\,\,\,\,\, & \mathrm{\Delta}\mathrm{v}& \mathrm{So.} & \mathrm{Cl.}\\
  & & & &\mathrm{(\,km}&\mathrm{s^{-1}\,)}& & \\
            \hline
            \noalign{\smallskip}

  1           &11\ 22\ 00.12 ,+42\ 54\ 25.2 & 17.76 &    15.75&  23230&   53& \sds & 0 \\
  2           &11\ 22\ 03.36 ,+43\ 11\ 41.6 & 18.53 &    18.25& 180587&  338& \sds & 0 \\
  3           &11\ 22\ 04.44 ,+43\ 01\ 35.4 & 18.57 &    16.87&  37090&   40& \sds & 0 \\
  4           &11\ 22\ 24.24 ,+42\ 53\ 08.9 & 18.66 &    17.46&  44576&   27& \sds & 0 \\
  5           &11\ 22\ 41.52 ,+43\ 00\ 41.4 & 18.47 &    17.17&  21546&   49& \sds & 0 \\
  6           &11\ 22\ 46.20 ,+42\ 57\ 37.4 & 19.28 &    17.28&  33915&   46& \sds & 0 \\
  7           &11\ 22\ 59.88 ,+43\ 06\ 43.6 & 18.48 &    17.10&  34993&   32& \sds & 0 \\
  8           &11\ 23\ 08.52 ,+43\ 10\ 41.9 & 18.93 &    17.18&  26827&   42& \sds & 0 \\
  9           &11\ 23\ 09.24 ,+42\ 55\ 38.3 & 19.10 &    17.39&  58448&   31& \sds & 2 \\
 10           &11\ 23\ 10.32 ,+42\ 53\ 06.2 & 18.49 &    17.10&  33987&   41& \sds & 0 \\
 11           &11\ 23\ 11.96 ,+42\ 56\ 08.7 & 20.40 &    19.20&  58737&   78& \tng & *2 \\
 12           &11\ 23\ 14.37 ,+42\ 57\ 28.6 & 21.00 &    19.34&  90603&   72& \tng & *0 \\
 13           &11\ 23\ 15.00 ,+43\ 12\ 08.6 & 19.24 &    17.56&  24005&   30& \sds & 0 \\
 14           &11\ 23\ 15.76 ,+42\ 58\ 06.7 & 23.01 &    21.11&  56786&  150& \tng & 2 \\
 15           &11\ 23\ 16.45 ,+42\ 53\ 50.3 & 20.20 &    18.05&  57516&   56& \tng & 2 \\
 16           &11\ 23\ 18.74 ,+42\ 57\ 30.1 & 20.38 &    18.68&  73676&   88& \tng & *0 \\
 17           &11\ 23\ 19.68 ,+43\ 18\ 44.6 & 21.96 &    18.84& 133977&   49& \sds & 0 \\
 18           &11\ 23\ 20.92 ,+42\ 56\ 28.3 & 21.40 &    19.08&  57994&  100& \tng & 2 \\
 19           &11\ 23\ 21.02 ,+42\ 57\ 12.5 & 19.58 &    17.91&  57551&   72& \tng & 2 \\
 20           &11\ 23\ 21.69 ,+42\ 55\ 25.9 & 20.22 &    17.83&  57982&   52& \tng & 2 \\
 21           &11\ 23\ 22.07 ,+42\ 54\ 58.6 & 21.64 &    19.22&  59445&   92& \tng & 2 \\
 22           &11\ 23\ 22.07 ,+43\ 04\ 44.2 & 20.81 &    18.64&  59551&   83& \tng & 1 \\
 23           &11\ 23\ 23.23 ,+42\ 54\ 27.7 & 21.22 &    19.03&  58431&   82& \tng & 2 \\
 24           &11\ 23\ 23.70 ,+42\ 56\ 13.9 & 20.29 &    17.88&  58813&   38& \tng & 2 \\
 25           &11\ 23\ 23.88 ,+42\ 58\ 25.7 & 23.49 &    19.78&  90451&  142& \tng & *0 \\
 26           &11\ 23\ 24.26 ,+43\ 09\ 38.1 & 20.57 &    18.12&  59406&  140& \tng & 1 \\
\textbf{27}   &11\ 23\ 24.64 ,+42\ 56\ 06.0 & 19.46 &    16.92&  57965&   38& \tns & 2 \\
 28           &11\ 23\ 25.78 ,+42\ 51\ 34.7 & 22.14 &    20.45&  58742&  120& \tng & 2 \\
\textbf{29}   &11\ 23\ 25.80 ,+42\ 54\ 17.6 & 19.04 &    16.49&  57482&   52& \sds & 2 \\
 30           &11\ 23\ 26.04 ,+43\ 05\ 10.5 & 20.18 &    17.86&  59031&   50& \tng & 1 \\
 31           &11\ 23\ 27.35 ,+43\ 00\ 00.9 & 20.28 &    18.34&  35290&   70& \tng & 0 \\
 32           &11\ 23\ 27.80 ,+42\ 59\ 35.0 & 18.51 &    17.22&  17883&   26& \tns & *0 \\
 33           &11\ 23\ 28.33 ,+43\ 05\ 48.1 & 21.23 &    19.27&  57149&  174& \tng & 1 \\
 34           &11\ 23\ 28.35 ,+42\ 53\ 49.1 & 20.77 &    18.52&  57489&   36& \tng & 2 \\
 35           &11\ 23\ 28.43 ,+43\ 05\ 29.9 & 20.49 &    18.22&  57988&   61& \tng & 1 \\
               
                        \noalign{\smallskip}			    
            \hline					    
            \noalign{\smallskip}			    
            \hline					    
         \end{array}
     $$ 
         \end{table}
\addtocounter{table}{-1}
\begin{table}[!ht]
          \caption[ ]{Continued.}
     $$ 
           \begin{array}{r c c r r r c r}
            \hline
            \noalign{\smallskip}
            \hline
            \noalign{\smallskip}

\mathrm{ID} & \mathrm{\alpha},\mathrm{\delta}\,(\mathrm{J}2000)  & B & R &\mathrm{v}\,\,\,\,\,\,\, & \mathrm{\Delta}\mathrm{v}& \mathrm{So.}& \mathrm{Cl.}\\
  & & & &\mathrm{(\,km}&\mathrm{s^{-1}\,)}& & \\

            \hline
            \noalign{\smallskip}

 36            &11\ 23\ 28.81 ,+42\ 52\ 09.7 & 21.81 &    20.33&  56626&   86& \tng & 2 \\
 37            &11\ 23\ 29.80 ,+42\ 53\ 31.6 & 21.82 &    19.72&  58741&   54& \tng & 2 \\
 38            &11\ 23\ 29.97 ,+42\ 57\ 51.3 & 19.90 &    17.63&  58621&  142& \tng & 2 \\
 39            &11\ 23\ 30.13 ,+43\ 06\ 58.6 & 21.41 &    18.79&  97901&  128& \tng & 0 \\
 40            &11\ 23\ 30.17 ,+43\ 02\ 18.2 & 20.45 &    18.67&  57717&  112& \tng & 1 \\
 41            &11\ 23\ 30.84 ,+42\ 54\ 55.8 & 19.71 &    17.30&  57368&   43& \sds & 2 \\
 42            &11\ 23\ 31.07 ,+42\ 59\ 27.6 & 21.74 &    19.63& 102577&  106& \tng & 0 \\
 43            &11\ 23\ 31.17 ,+42\ 53\ 53.3 & 19.30 &    18.16&  44948&   88& \tng & *0 \\
 44            &11\ 23\ 31.20 ,+42\ 54\ 45.8 & 19.84 &    17.54&  59205&   60& \tng & 2 \\
 45            &11\ 23\ 31.32 ,+42\ 54\ 26.3 & 21.30 &    19.31&  59217&   84& \tng & 2 \\
 46            &11\ 23\ 32.10 ,+43\ 08\ 42.0 & 20.75 &    18.41&  57592&   81& \tng & 1 \\
 47            &11\ 23\ 32.40 ,+43\ 09\ 06.8 & 22.03 &    19.59&  58446&   76& \tng & 1 \\
 48            &11\ 23\ 32.68 ,+43\ 07\ 56.0 & 20.83 &    18.79&  58949&  148& \tng & 1 \\
 49            &11\ 23\ 32.75 ,+42\ 57\ 31.8 & 19.75 &    17.60&  57246&   68& \tng & 2 \\
 50            &11\ 23\ 33.08 ,+43\ 06\ 32.2 & 20.18 &    17.86&  59329&  150& \tng & 1 \\
 51            &11\ 23\ 33.72 ,+43\ 14\ 47.4 & 19.67 &    17.25&  59070&   49& \sds & 1 \\
 52            &11\ 23\ 33.84 ,+43\ 08\ 53.3 & 20.96 &    18.62&  84868&  112& \tng & *0 \\
 53            &11\ 23\ 34.15 ,+43\ 08\ 12.9 & 20.35 &    18.07&  57839&  128& \tng & 1 \\
 54            &11\ 23\ 34.67 ,+43\ 04\ 12.1 & 20.50 &    18.24&  58620&   64& \tng & 1 \\
 55            &11\ 23\ 34.82 ,+43\ 01\ 56.9 & 20.14 &    19.22&  23909&   87& \tng & *0 \\
 \textbf{56}   &11\ 23\ 35.26 ,+43\ 08\ 31.7 & 18.74 &    16.22&  58353&   96& \tng & 1 \\
 57            &11\ 23\ 35.46 ,+43\ 04\ 57.7 & 19.57 &    16.97&  58717&   46& \tns & 1 \\
 58            &11\ 23\ 35.48 ,+43\ 05\ 13.7 & 19.84 &    17.23&  58230&  100& \tng & 1 \\
 59            &11\ 23\ 35.52 ,+42\ 53\ 43.8 & 19.23 &    17.46&  58805&   43& \sds & 2 \\
 60            &11\ 23\ 35.60 ,+43\ 05\ 06.1 & 20.79 &    18.93&  58415&  180& \tng & 1 \\
 61            &11\ 23\ 35.77 ,+43\ 11\ 19.2 & 20.99 &    18.80&  88635&  156& \tng & *0 \\
 62            &11\ 23\ 36.00 ,+43\ 16\ 06.3 & 21.02 &    18.99&  53996&   94& \tng & 0 \\
 63            &11\ 23\ 36.34 ,+43\ 07\ 11.1 & 20.99 &    18.77&  58358&  204& \tng & 1 \\
 64            &11\ 23\ 36.56 ,+42\ 57\ 20.5 & 19.67 &    17.58&  56715&   78& \tng & 2 \\
 65            &11\ 23\ 36.68 ,+43\ 09\ 01.9 & 20.38 &    18.00&  60953&  116& \tng & 0 \\
 66            &11\ 23\ 36.77 ,+42\ 59\ 13.5 & 21.03 &    18.88&  57097&   84& \tng & 2 \\
 67            &11\ 23\ 36.77 ,+43\ 11\ 44.5 & 20.99 &    19.00&  59749&   92& \tng & 1 \\
 68            &11\ 23\ 36.88 ,+43\ 04\ 56.3 & 22.13 &    19.75&  59656&  192& \tng & 1 \\
 69            &11\ 23\ 36.91 ,+43\ 15\ 37.3 & 22.17 &    20.82& 153440&  130& \tng & *0 \\
 70            &11\ 23\ 37.16 ,+42\ 53\ 46.7 & 20.28 &    18.81&  56870&   74& \tng & *2 \\

               \noalign{\smallskip}			    
            \hline					    
            \noalign{\smallskip}			    
            \hline					    
         \end{array}
     $$ 
         \end{table}
\addtocounter{table}{-1}
\begin{table}[!ht]
          \caption[ ]{Continued.}
     $$ 
           \begin{array}{r c c r r r c r}
            \hline
            \noalign{\smallskip}
            \hline
            \noalign{\smallskip}

\mathrm{ID} & \mathrm{\alpha},\mathrm{\delta}\,(\mathrm{J}2000)  & B & R &\mathrm{v}\,\,\,\,\,\,\, & \mathrm{\Delta}\mathrm{v}& \mathrm{So.}& \mathrm{Cl.}\\
  & & & &\mathrm{(\,km}&\mathrm{s^{-1}\,)}& & \\

            \hline
            \noalign{\smallskip}

 71            &11\ 23\ 37.17 ,+43\ 11\ 31.4 & 20.36 &    18.15&  58635&  142& \tng & 1 \\
 72            &11\ 23\ 37.28 ,+43\ 01\ 16.6 & 21.38 &    19.31&  59878&   70& \tng & 1 \\
 73            &11\ 23\ 37.34 ,+43\ 09\ 10.2 & 21.47 &    19.01& 100284&  258& \tng & 0 \\
 74            &11\ 23\ 37.53 ,+43\ 04\ 19.9 & 21.90 &    19.79&  57709&  118& \tng & 1 \\
 \textbf{75}   &11\ 23\ 37.69 ,+43\ 03\ 28.2 & 18.82 &    16.12&  58817&   54& \tng & 1 \\
 76            &11\ 23\ 37.83 ,+43\ 03\ 10.8 & 19.76 &    17.11&  56550&   78& \tng & 1 \\
 77            &11\ 23\ 38.13 ,+43\ 09\ 43.9 & 20.84 &    19.42&  60307&  194& \tng & *1 \\
 78            &11\ 23\ 38.22 ,+43\ 03\ 38.6 & 20.23 &    18.01&  59542&   92& \tng & 1 \\
 79            &11\ 23\ 38.32 ,+43\ 09\ 39.9 & 21.24 &    19.26&  58342&  140& \tng & 1 \\
 80            &11\ 23\ 38.35 ,+43\ 07\ 42.0 & 20.78 &    18.53&  57397&   53& \tng & 1 \\
 81            &11\ 23\ 38.37 ,+43\ 09\ 55.0 & 20.80 &    19.91&  59828&  148& \tng & 1 \\
 82            &11\ 23\ 38.46 ,+43\ 07\ 48.2 & 20.90 &    19.57&  58605&  200& \tng & 1 \\
 83            &11\ 23\ 38.63 ,+43\ 01\ 49.1 & 21.23 &    18.87&  58509&   80& \tng & 1 \\
 84            &11\ 23\ 38.85 ,+43\ 04\ 36.8 & 20.57 &    19.33&  55562&  134& \tng & *1 \\
 85            &11\ 23\ 39.15 ,+43\ 02\ 35.5 & 20.77 &    18.73&  57478&   78& \tng & 1 \\
 86            &11\ 23\ 39.54 ,+43\ 10\ 07.4 & 22.07 &    19.63&  58711&  146& \tng & 1 \\
 87            &11\ 23\ 40.08 ,+43\ 10\ 36.4 & 20.74 &    18.37&  57351&  160& \tng & 1 \\
 88            &11\ 23\ 40.47 ,+43\ 09\ 18.7 & 21.41 &    18.97&  59078&   94& \tng & 1 \\
 89            &11\ 23\ 40.50 ,+43\ 02\ 59.3 & 20.92 &    18.79&  56851&   82& \tng & 1 \\
 90            &11\ 23\ 40.87 ,+42\ 53\ 37.4 & 17.00 &    15.19&  23141&   37& \tns & 0 \\
 91            &11\ 23\ 41.42 ,+42\ 52\ 33.5 & 20.59 &    19.47&  58938&  118& \tng & *2 \\
 92            &11\ 23\ 41.92 ,+43\ 10\ 51.7 & 19.76 &    17.41&  57628&  138& \tng & 1 \\
 93            &11\ 23\ 42.04 ,+43\ 01\ 03.9 & 21.19 &    19.61&  56216&  120& \tng & 1 \\
 94            &11\ 23\ 42.05 ,+43\ 12\ 17.8 & 19.45 &    17.16&  57828&   42& \tns & 1 \\
 95            &11\ 23\ 42.60 ,+43\ 15\ 17.3 & 20.91 &    18.65&  59502&  120& \tng & 1 \\
 96            &11\ 23\ 42.85 ,+43\ 09\ 29.3 & 22.03 &    19.57&  57407&  144& \tng & 1 \\
 97            &11\ 23\ 42.89 ,+43\ 12\ 24.4 & 21.43 &    18.91&  90672&   92& \tng & 0 \\
 98            &11\ 23\ 42.94 ,+42\ 56\ 16.2 & 18.87 &    17.55&  23440&   94& \tng & *0 \\
 99            &11\ 23\ 43.03 ,+43\ 10\ 09.4 & 21.59 &    19.79&  90196&  186& \tng & 0 \\
100            &11\ 23\ 44.16 ,+43\ 17\ 44.5 & 18.97 &    17.96&  59099&   29& \sds & 1 \\
101            &11\ 23\ 44.19 ,+42\ 53\ 30.4 & 21.03 &    18.45&  57649&   46& \tng & 2 \\
102            &11\ 23\ 44.90 ,+43\ 14\ 02.4 & 21.25 &    19.06&  59238&  186& \tng & 1 \\
103            &11\ 23\ 45.26 ,+43\ 08\ 14.8 & 21.35 &    19.93&  56860&  110& \tng & 1 \\
104            &11\ 23\ 45.29 ,+43\ 12\ 38.4 & 21.18 &    18.88&  98761&   94& \tng & 0 \\
105            &11\ 23\ 46.58 ,+42\ 59\ 57.6 & 20.57 &    18.78&  56478&   76& \tng & 1 \\

               \noalign{\smallskip}			    
            \hline					    
            \noalign{\smallskip}			    
            \hline					    
         \end{array}
     $$ 
         \end{table}
\addtocounter{table}{-1}
\begin{table}[!ht]
          \caption[ ]{Continued.}
     $$ 
           \begin{array}{r c c r r r c r}
            \hline
            \noalign{\smallskip}
            \hline
            \noalign{\smallskip}

\mathrm{ID} & \mathrm{\alpha},\mathrm{\delta}\,(\mathrm{J}2000)  & B & R &\mathrm{v}\,\,\,\,\,\,\, & \mathrm{\Delta}\mathrm{v}& \mathrm{So.}& \mathrm{Cl.}\\
  & & & &\mathrm{(\,km}&\mathrm{s^{-1}\,)}& & \\

            \hline
            \noalign{\smallskip}

106            &11\ 23\ 48.35 ,+43\ 11\ 11.6 & 20.54 &    18.95&  58801&   84& \tng & *1 \\
107            &11\ 23\ 49.20 ,+43\ 08\ 07.9 & 20.39 &    18.30&  57904&  118& \tng & 1 \\
108            &11\ 23\ 49.42 ,+43\ 00\ 17.4 & 21.08 &    18.78&  58494&  116& \tng & 1 \\
109            &11\ 23\ 49.54 ,+43\ 07\ 59.1 & 22.51 &    20.07&  84937&  196& \tng & 0 \\
110            &11\ 23\ 49.83 ,+43\ 03\ 52.8 & 21.37 &    18.79&  91883&  118& \tng & 0 \\
\textbf{111}   &11\ 23\ 49.92 ,+43\ 05\ 44.7 & 19.21 &    16.72&  56134&   66& \tng & 1 \\
112            &11\ 23\ 50.86 ,+43\ 13\ 22.2 & 21.25 &    19.06&  59144&  104& \tng & 1 \\
113            &11\ 23\ 51.09 ,+43\ 12\ 01.0 & 21.51 &    19.94&  61506&  118& \tng & 0 \\
114            &11\ 23\ 51.14 ,+43\ 00\ 09.2 & 20.62 &    19.24&  48223&  126& \tng & *0 \\
115            &11\ 23\ 51.15 ,+43\ 12\ 06.5 & 20.98 &    19.59&  78030&  138& \tng & *0 \\
116            &11\ 23\ 51.87 ,+43\ 11\ 54.8 & 20.96 &    18.85& 152164&  138& \tng & 0 \\
117            &11\ 23\ 52.35 ,+43\ 12\ 45.5 & 21.11 &    19.50&  55225&   94& \tng & 0 \\
118            &11\ 23\ 52.44 ,+42\ 54\ 36.4 & 18.55 &    17.45&  23511&   28& \sds & 0 \\
119            &11\ 23\ 52.44 ,+43\ 03\ 28.1 & 19.62 &    17.15&  58336&   48& \sds & 1 \\
120            &11\ 23\ 52.47 ,+42\ 58\ 19.6 & 21.25 &    19.12&  58639&   86& \tng & 1 \\
121            &11\ 23\ 52.51 ,+43\ 14\ 50.0 & 20.93 &    18.83&  59284&  104& \tng & 1 \\
122            &11\ 23\ 52.59 ,+43\ 06\ 05.3 & 22.05 &    19.03&  84545&  166& \tng & 0 \\
123            &11\ 23\ 53.52 ,+43\ 06\ 13.3 & 19.76 &    17.30&  59022&   47& \sds & 1 \\
124            &11\ 23\ 54.07 ,+43\ 15\ 07.2 & 20.21 &    18.63&  21528&   86& \tng & 0 \\
125            &11\ 23\ 54.10 ,+43\ 11\ 24.4 & 20.63 &    18.97&  77906&   94& \tng & *0 \\
126            &11\ 23\ 54.21 ,+43\ 14\ 17.2 & 21.87 &    19.94& 106774&   98& \tng & 0 \\
127            &11\ 23\ 54.31 ,+42\ 58\ 28.1 & 20.48 &    18.59&  87022&   90& \tng & 0 \\
128            &11\ 23\ 54.60 ,+43\ 04\ 30.0 & 17.76 &    16.07&  21275&   45& \sds & 0 \\
129            &11\ 23\ 55.13 ,+43\ 14\ 26.0 & 19.61 &    18.90&  24046&   30& \tng & *0 \\
130            &11\ 23\ 56.24 ,+42\ 59\ 40.4 & 19.98 &    17.58&  58805&   68& \tng & 1 \\
131            &11\ 23\ 56.76 ,+42\ 58\ 06.2 & 18.00 &    16.57&  23303&   31& \sds & 0 \\
132            &11\ 23\ 57.23 ,+43\ 09\ 56.6 & 21.71 &    20.56&  58966&  130& \tng & *1 \\
133            &11\ 23\ 57.48 ,+43\ 13\ 13.8 & 21.14 &    18.78& 331328&  450& \sds & 0 \\
134            &11\ 23\ 57.57 ,+43\ 13\ 01.2 & 19.36 &    17.22&  54792&   80& \tng & 0 \\
135            &11\ 23\ 57.81 ,+43\ 10\ 41.5 & 21.44 &    19.24&  58346&  156& \tng & 1 \\
136            &11\ 24\ 01.38 ,+42\ 58\ 55.5 & 20.18 &    18.45&  48099&   74& \tng & 0 \\
137            &11\ 24\ 01.72 ,+43\ 15\ 50.5 & 21.28 &    20.33&  78070&   71& \tng & *0 \\
138            &11\ 24\ 01.75 ,+43\ 00\ 42.0 & 21.17 &    19.56&  57426&  144& \tng & *1 \\
139            &11\ 24\ 01.80 ,+43\ 08\ 57.5 & 17.77 &    16.31&  21326&   56& \sds & 0 \\
140            &11\ 24\ 02.20 ,+43\ 02\ 44.4 & 22.03 &    19.78&  56604&  212& \tng & 1 \\
141            &11\ 24\ 03.24 ,+43\ 13\ 30.7 & 19.66 &    18.04&  96076&   37& \sds & 0 \\
142            &11\ 24\ 03.60 ,+42\ 56\ 11.4 & 19.51 &    17.03&  58204&   46& \sds & 1 \\
143            &11\ 24\ 04.83 ,+43\ 13\ 48.0 & 17.54 &    15.48&  21158&  112& \tng & 0 \\
144            &11\ 24\ 07.19 ,+43\ 13\ 32.7 & 20.41 &    18.03&  55467&  102& \tng & 0 \\
145            &11\ 24\ 31.68 ,+42\ 56\ 49.6 & 21.16 &    18.28&  92075&   75& \sds & 0 \\

               \noalign{\smallskip}			    
            \hline					    
            \noalign{\smallskip}			    
            \hline					    
         \end{array}
     $$ 
         \end{table}


\section{Analysis and Results}
\label{anal}

\subsection{Member selection}
\label{memb}

\begin{figure}
\centering
\resizebox{\hsize}{!}{\includegraphics{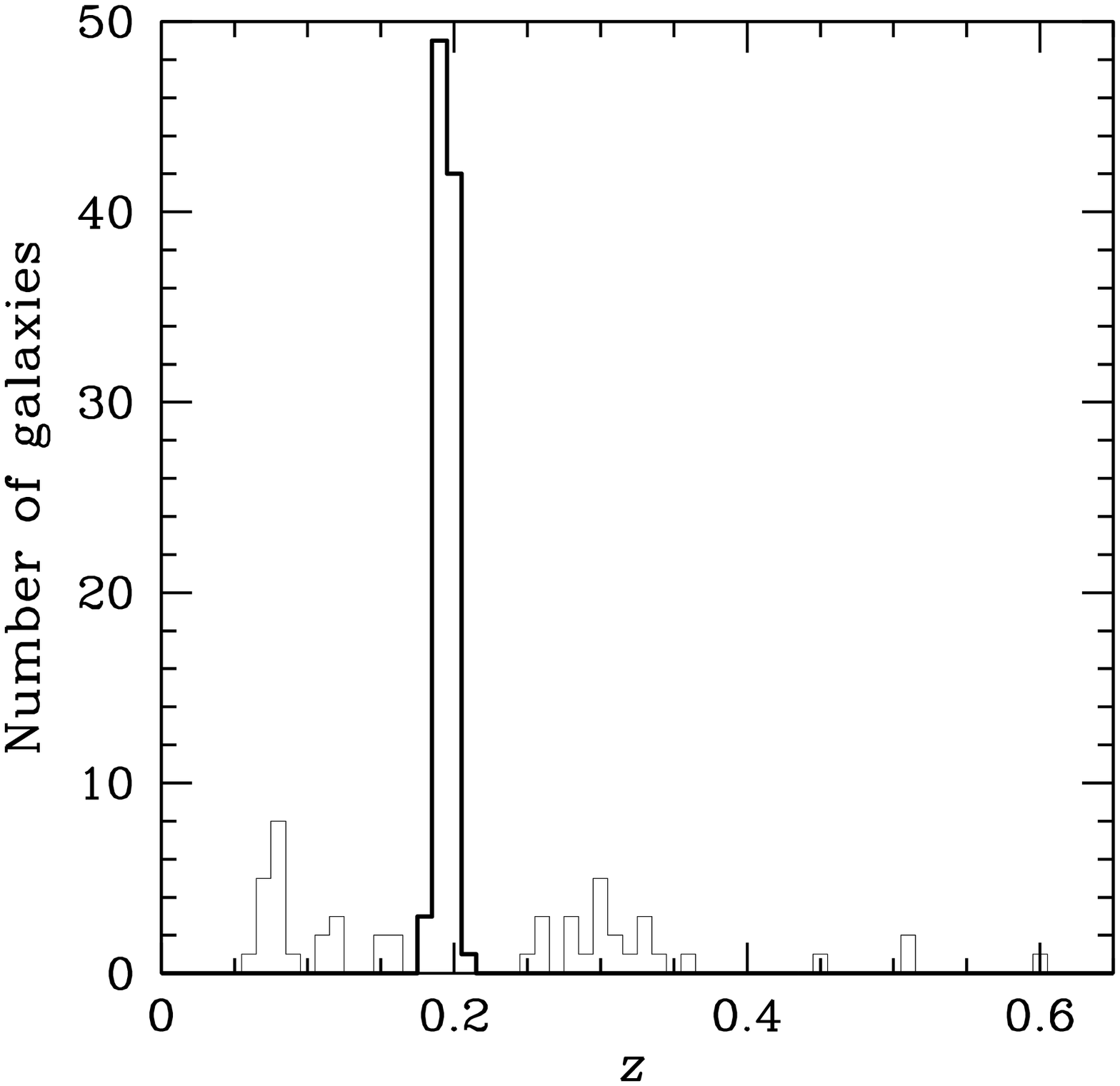}}
\caption
{Redshift galaxy distribution. The solid line histogram refers to
  the 95 galaxies assigned to the A1237+A1240 complex according to the DEDICA
  reconstruction method.}
\label{fighisto}
\end{figure}

To select cluster members out of 145 galaxies having redshifts, we
follow a two steps procedure. First, we perform the 1D
adaptive--kernel method (hereafter DEDICA, Pisani \cite{pis93} and
\cite{pis96}; see also Fadda et al. \cite{fad96}; Girardi et
al. \cite{gir96}). We search for significant peaks in the velocity
distribution at $>$99\% c.l.. This procedure detects A1237+A1240 as a
peak at $z\sim0.196$ populated by 95 galaxies considered as candidate
cluster members (see Fig.~\ref{fighisto}). Out of 50 non members, 24
and 26 are foreground and background galaxies, respectively. 

All the galaxies assigned to the cluster peak are analyzed in the second
step which uses the combination of position and velocity information:
the ``shifting gapper'' method by Fadda et al. (\cite{fad96}).  This
procedure rejects galaxies that are too far in velocity from the main
body of galaxies within a fixed bin that shifts along the distance
from the cluster center.  The procedure is iterated until the number
of cluster members converges to a stable value.  Following Fadda et
al. (\cite{fad96}) we use a gap of $1000$ \ks -- in the cluster
rest--frame -- and a bin of 0.6 \hh, or large enough to include 15
galaxies. 

\begin{figure}
\centering 
\resizebox{\hsize}{!}{\includegraphics{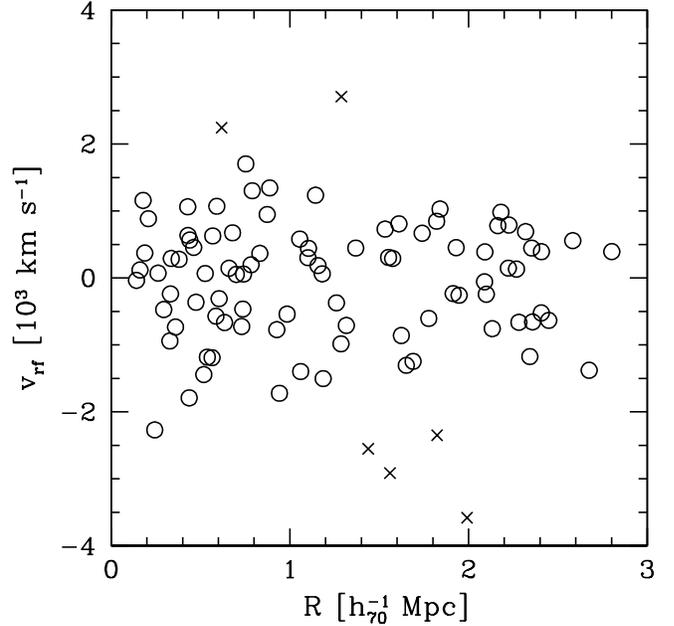}}
\caption
{ Rest--frame velocity vs. projected clustercentric
  distance for the 95 galaxies in the main peak (Fig.~\ref{fighisto})
  showing galaxies detected as interlopers by our ``shifting gapper''
  procedure (crosses). Circles indicate the 89 cluster members. 
  }
\label{figvd}
\end{figure}

The choice of the center of A1240 is not obvious. No evident dominant
galaxy is present, rather there are some luminous galaxies. In
particular, the two brightest ones, ID.~75 and ID.~56, lie in the
southern and northern region of A1240, respectively, and show
comparable luminosity (hereafter BCGS and BCGN). The third one is
located in the eastern region, but is $\gtrsim 0.5$ $R$--magnitudes
fainter than BCGS and BCGN (ID.~111, hereafter BCGE). As for the
cluster center, hereafter we assume the position of the centroid of
the X--ray emission as recovered by David et al. (\cite{dav99})
[R.A.=$11^{\mathrm{h}}23^{\mathrm{m}}37\dotsec6$, Dec.=$+43\degree
05\arcmm 51\arcs$ (J2000.0)] which lies between the two dominant
galaxies. After the ``shifting gapper'' procedure we obtain a sample
of 89 fiducial cluster members (see Fig.~\ref{figvd}).

We also check the result of alternative member selection
  procedures.  We apply the ``shifting gapper'' procedure adopting
  as cluster center the brightest cluster galaxy (BCGS). We find a
  cluster sample of 89 galaxies, 88 of which are in common with our
  above sample.  In order to analyze the effect of a fully alternative
  selection procedure we also apply the classical 3--$\sigma$
  clipping procedure by Yahil \& Vidal (1977) on the whole sample of
  145 galaxies after a very rough cut in the velocity space,
  i.e. rejecting galaxies with velocities differing by more than 8000
  \ks from the mean velocity. This classical procedure leads to a
  sample of 90 galaxies, 89 of which forms our adopted sample.  In
  conclusion, the sample of member galaxies we adopt in this work is
  quite robust against the member selection procedure.

\begin{figure}
\hspace{1.5cm}
\includegraphics[width=8cm]{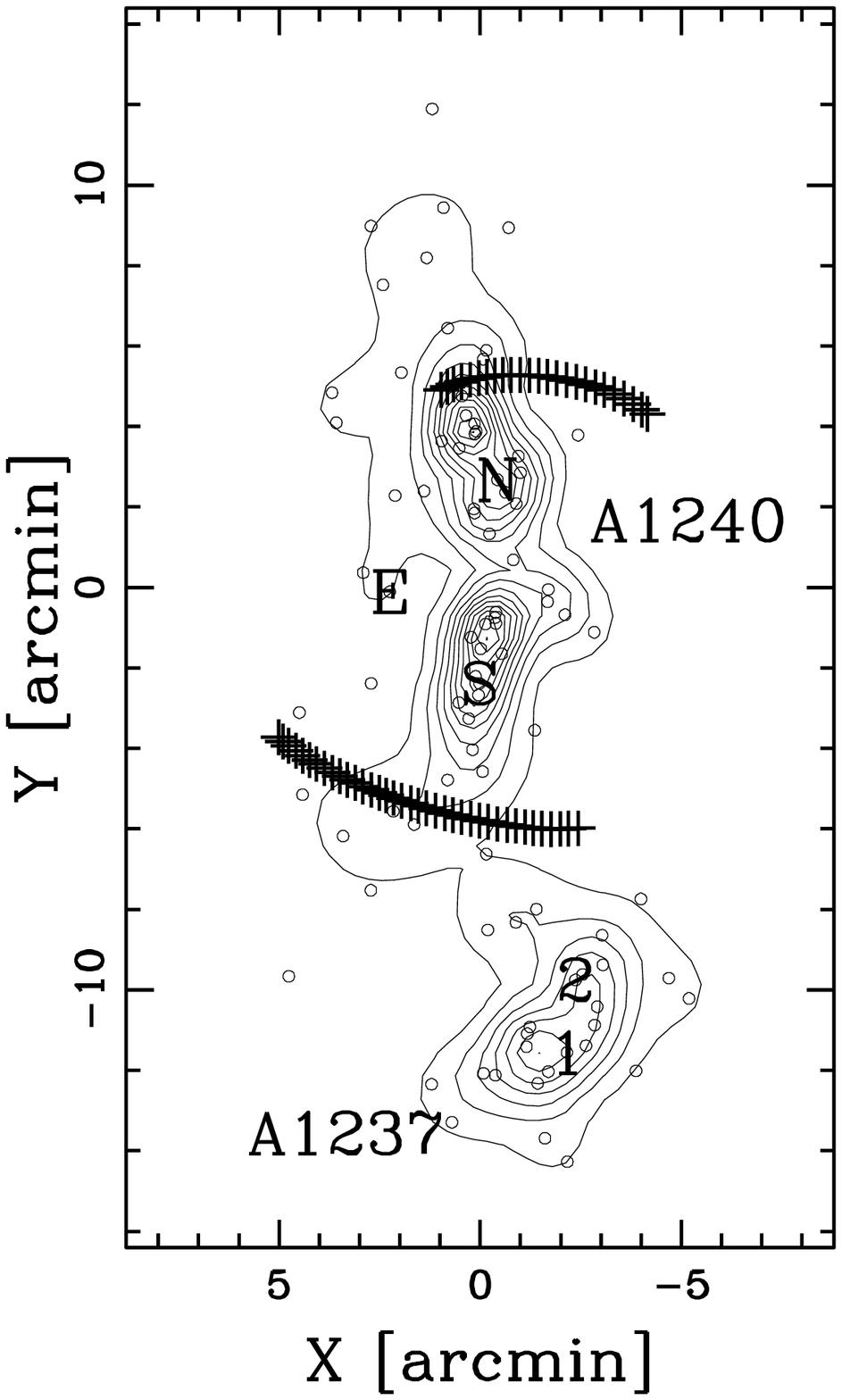}
\caption
{Spatial distribution on the sky and relative isodensity contour map
  of cluster members (A1237+A1240), obtained with the DEDICA method.
  The positions of the brightest galaxies are indicated (BCGN, BCGS,
  BCGE for A1240 and BCG1 and BCG2 for A1237). The plot is centered on
  the cluster center defined in this paper as the X--ray center.
  The two relics  are indicated in a schematic way.}
\label{figk2z}
\end{figure}

The galaxy distribution analyzed through the 2D DEDICA method clearly
shows the presence of a southern external clump (see
Fig.~\ref{figk2z}, see also Sect.~\ref{2D}).  Gal et
al. (\cite{gal03}) recovered a cluster in the same position from the
digitized Second Palomar Observatory Sky Survey. We identify this
galaxy clump with A1237 which is likely to have a cluster redshift
similar to that of A1240 (cf. the magnitudes of their 10th--ranked
cluster galaxies, Abell et al. \cite{abe89}). Notice that the center
reported by Abell et al. is quite imprecise and lies on the southern
border of the galaxy concentration we detect.

\begin{figure}
\centering
\resizebox{\hsize}{!}{\includegraphics{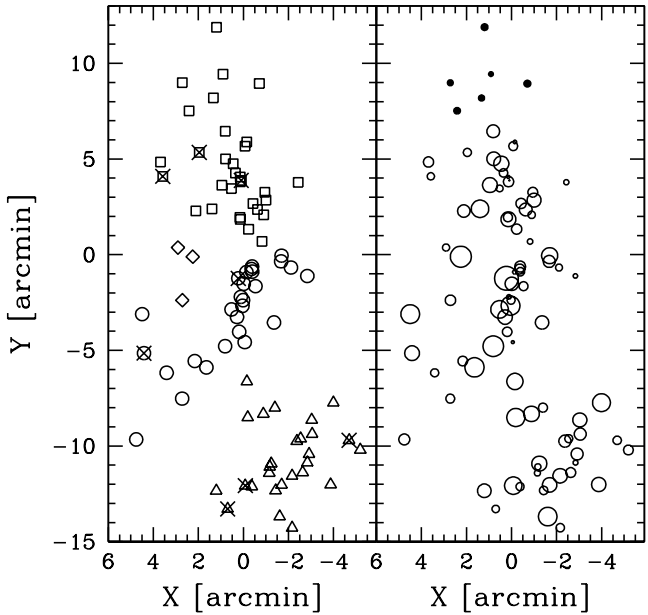}}
\caption
{Spatial distribution on the sky of 89 cluster members.  {\em Left
    panel}: open symbols indicate cluster 
    members. In particular, squares, circles, rotated squares and 
    triangles indicate
    A1240N, A1240S, A1240E and A1237. The crosses
    indicate emission line galaxies. {\em Right panel}: the larger the
    symbol, the smaller is the radial velocity. The six close
    northern circles indicate the high velocity galaxies we consider
    responsible of the apparent large global mean velocity of A1240N
    galaxies (see Sect.~\ref{int} and Fig.~\ref{figprofcori} in the
    following, too).}
\label{figvel}
\end{figure}

 We use the 2D DEDICA results, i.e. the peaks detected in the 2D
  galaxy distribution, to assign galaxies to different subclumps.  The
  2D DEDICA algorithm detects nine peaks, four of which are more
  significant than the $99\%$ c.l..  The southern three peaks, only one of
  which is very significant, are assigned to A1237 (for a
  total of 27 members). The six northern peaks, three of which  are
  very significant, are assigned to A1240 (for a total of 62 members).
This assignment is shown in Fig.~\ref{figvel} -- left panel.

As for A1240, the 2D DEDICA algorithm shows a clear bimodal structure
(see Fig.~\ref{figk2z}) along the North-South direction.  This
bimodality is also shown in our analysis of photometric ``likely''
members in Sect.~\ref{2D} and corresponds to the elongated hot gas
distribution shown by the previous analyses of ROSAT data (David et
al. \cite{dav99}; Bonafede et al. \cite{bon09}) and our analysis of
Chandra data (see Sect.~\ref{xray}). Therefore we decide to consider:
a southern structure -- hereafter A1240S -- associated to the southern
peak of A1240 (the most significant in the whole DEDICA analysis); a
northern structure -- hereafter a1240N -- associated to the four
northern peaks (two of which are very significant).  In this way we
assign 32 (27) members to A1240N (A1240S). A1240S and A1240N host the
brightest and the second brightest galaxies, BCGS and BCGN,
respectively.  We consider separately three galaxies belonging to a
minor, eastern peak (hereafter A1240E) since their assignation to
A1240N or A1240S is not obvious. A1240E hosts the BCGE. The
assignment of galaxies within A1240 is summarized as follow: we
assign to A1240S the galaxies belonging to the southern peak
(detected with a $>99.99\%$ c.l.); to A1240N the galaxies belonging
to the four, northern peaks (two of which are detected with a $>
99.8\%$ c.l.); to a1240E the galaxies belonging to the eastern
peak.

\subsection{The A1237+A1240 complex in the velocity space}
\label{vel}

According to the 1D Kolmogorov--Smirnov test (hereafter 1DKS--test;
see, e.g., Press et al \cite{pre92}), there is no significant
difference between the velocity distributions of A1237 and A1240 (see
also Fig.~\ref{figvel} -- right panel). This result suggests us to
investigate the global velocity distribution of the complex.

\begin{figure}
\centering
\resizebox{\hsize}{!}{\includegraphics{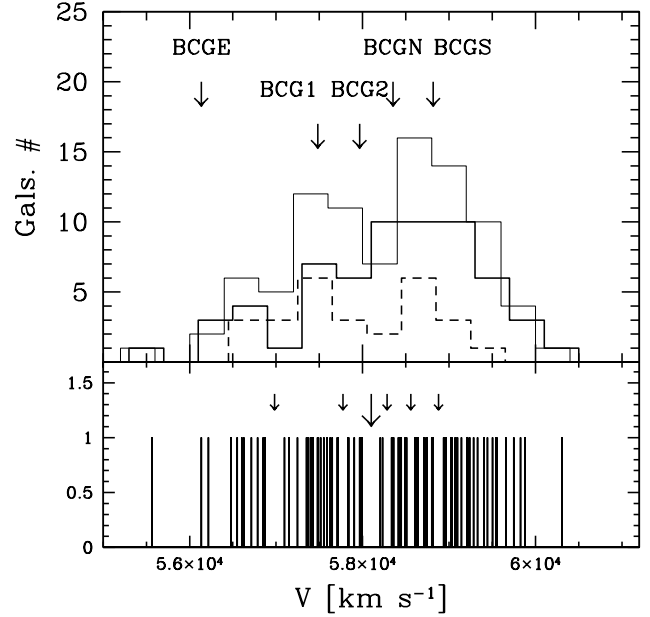}}
\caption
{{\em Upper panel}: velocity histogram for the whole cluster sample
  (thin line), A1240 (solid line) and A1237 (dashed line). The
  velocities of the three brightest galaxies of A1240 are indicated
(BCGN, BCGS and BCGE), as well as the two brightest galaxies of A1237
  (BCG1 and BCG2).  {\em Lower panel}: stripe density plot where
  the arrows indicate the positions of the significant gaps.
  The main and minor weighted gaps are indicated
  by big and small arrows, respectively.}
\label{figstrip}
\end{figure}

We analyze the velocity distribution of cluster galaxies (see
Fig.~\ref{figstrip}) using a few tests where the null hypothesis is
that the velocity distribution is a single Gaussian. We estimate three
shape estimators: the kurtosis, the skewness, and the scaled tail
index (see, e.g., Beers et al.~\cite{bee91}). According to the value
of the skewness (-0.379) the velocity distribution is marginally
asymmetric differing from a Gaussian at the $90-95\%$ c.l. (see
Table~2 of Bird \& Beers \cite{bir93}). We also analyze the presence
of ``weighted gaps'' in the velocity distribution. A weighted gap in
the space of the ordered velocities is defined as the difference
between two contiguous velocities, weighted by the location of these
velocities with respect to the middle of the data (Wainer and
Schacht~\cite{wai78}; Beers et al.~\cite{bee91}). We detect a strongly
significant gap (at the $>99.9\%$ c.l.) and five minor gaps (at the
$\gtrsim 98\%$ c.l.), see Fig.~\ref{figstrip} -- lower panel. The most
important gap, very significant since it is located in the central
region of the velocity distribution, separates the cluster into two
subgroups of 37 and 52 galaxies. When comparing the 2D galaxy
distributions of these subgroups we find no difference according to
the 2D Kolmogorov--Smirnov test (Fasano \& Franceschini \cite{fas87}).

We also perform the 2D and 3D Kaye's mixture model (KMM) test (as
implemented by Ashman et al. \cite{ash94}) and compare the results to
check the effect of the addition of the velocity information. The KMM
algorithm fits a user--specified number of Gaussian distributions to a
dataset and assesses the improvement of that fit over a single
Gaussian and give an assignment of objects into groups. We use the
A1237+A1240 galaxy assignment found by the 2D DEDICA analysis to
determine the first guess when fitting two groups. The 2D KMM
algorithm fits a two--group partition, at the $>99.9\%$ c.l. according
to the likelihood ratio test, leading to two groups of 66 and 23
galaxies. The addition of the velocity information in the KMM
algorithm leads to the same group partition.

Finally, we combine galaxy velocity and position information to
compute the $\Delta$--statistics devised by Dressler \& Shectman
(\cite{dre88}). This test is sensitive to spatially compact subsystems
that have either an average velocity that differs from the cluster
mean, or a velocity dispersion that differs from the global one, or
both. We find no significant substructure.

We conclude that, although the velocity distribution shows evidence of
a complex structure, A1237 and A1240 are so similar in the velocity
space that the velocity information is not useful to improve the
galaxy assignment recovered from the 2D analysis (see Sect.~\ref{memb}).

\subsection{Global Kinematical properties}
\label{kin}

As for the whole cluster complex, by applying the biweight estimator to
the 89 members (Beers et al. \cite{bee90}), we
compute a mean redshift of $\left<z\right>=0.1937\pm$ 0.0003, i.e.
$\left<\rm{v}\right>=(58273\pm 89)$ \kss.  We estimate the LOS velocity
dispersion, $\sigma_{\rm V}$, by using the biweight estimator and
applying the cosmological correction and the standard correction for
velocity errors (Danese et al. \cite{dan80}). We obtain $\sigma_{\rm
V}=842_{-55}^{+63}$ \kss, where errors are estimated through a
bootstrap technique.

\begin{figure}
\centering 
\resizebox{\hsize}{!}{\includegraphics{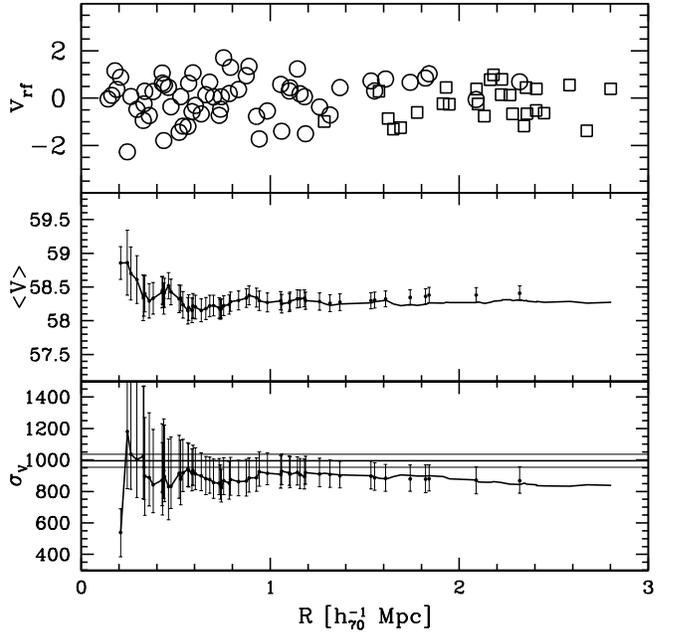}}
\caption
{ {\em Upper panel}: rest--frame velocity (in units of $10^3$ \kss)
  vs. projected clustercentric distance for the 89 cluster galaxies.
  Circles and squares indicate members of A1240 and A1237.  {Central
    and lower panels}: integral profiles of mean LOS velocity (in
  units of $10^3$ \kss) and LOS velocity dispersion (in units of \kss)
  for the 62 members of A1240 (dots with the $68\%$ c.l. errorbar) and
  for all the 89 cluster members (A1237+A1240, solid line).  The mean
  and dispersion at a given (projected) radius from the
  cluster--center is estimated by considering all galaxies within that
  radius -- the first value computed on the five galaxies closest to
  the center. In the lower panel, the horizontal lines represent the
  X--ray temperature with the respective $68\%$ errors transformed in
  $\sigma_{\rm V}$ assuming the density--energy equipartition between
  gas and galaxies, i.e. $\beta_{\rm spec}=1$ where $\beta_{\rm
    spec}=\sigma_{\rm V}^2/(kT/\mu m_{\rm p})$ with $\mu=0.58$ the
  mean molecular weight and $m_{\rm p}$ the proton mass. 
In all the panels the center is the X--ray center of A1240.}
\label{figprof}
\end{figure}

The results obtained for the 62 members of A1240 are:
$\left<z\right>=0.1948\pm$ 0.0004, i.e.
$\left<\rm{v}\right>=(58408\pm 111)$ \ks, and $\sigma_{\rm
  V}=870_{-79}^{+91}$ \kss.  To evaluate the robustness of 
$\sigma_{\rm V}$ of A1240 we analyze the velocity dispersion profile
(Fig.~\ref{figprof}).  The integral profile is roughly flat in the
external cluster regions suggesting that the value of $\sigma_{\rm V}$
for A1240 is quite robust. Figure~\ref{figprof} also shows that the
$\left<\rm{v}\right>$ and $\sigma_{\rm V}$ profiles are not disturbed
by the presence of A1237.

Table~\ref{tabv} lists the number of the member galaxies $N_{\rm g}$
and the main kinematical properties of A1237 and A1240.

\begin{table*}
        \caption[]{Global properties of galaxy systems.}
         \label{tabv}
            $$
         \begin{array}{l c c c c c c}
            \hline
            \noalign{\smallskip}
            \hline
            \noalign{\smallskip}
\mathrm{System} & N_{\rm g} & \alpha,\,\delta&\mathrm{<v>}&\sigma_{\rm V}&R_{\rm vir}&M(<R_{\rm vir})\\
& &({\rm J}2000)&\mathrm{km\ s^{-1}}&\mathrm{km\ s^{-1}}&h_{70}^{-1}{\rm Mpc}&h_{70}^{-1}10^{14}\mathrm{M}_{\odot}\\
         \hline
         \noalign{\smallskip}
\mathrm{A1237} &27&112329.5+425419&58021\pm145&738_{-54}^{+82} &1.6&6\pm2\\
\mathrm{A1240} &62&112337.6+430551&58408\pm111&870_{-79}^{+91}&1.9-2.4^{\mathrm{b}}&9-19^{\mathrm{b}}\\
\mathrm{A1240N^{\mathrm{a}}}&32&112335.2+430832&\sim 58353&709_{-83}^{+88}&1.6&5\pm2\\
\mathrm{A1240S^{\mathrm{a}}}&27&112337.7+430328&\sim 58817&991_{-99}^{+149}&2.2&14\pm5\\
              \noalign{\smallskip}
            \hline
            \noalign{\smallskip}
            \hline
         \end{array}
$$
\begin{list}{}{}  
\item[$^{\mathrm{a}}$] {\bf As center and mean velocity
of this clump we list the position and velocity} of
the respective brightest galaxy. 
\item[$^{\mathrm{b}}$] The lower limit comes from the observed
$\sigma_{\rm V}$. The upper limit is obtained when considering the bimodal structure of the cluster (see text).
\end{list}
         \end{table*}

\begin{figure}
\centering 
\resizebox{\hsize}{!}{\includegraphics{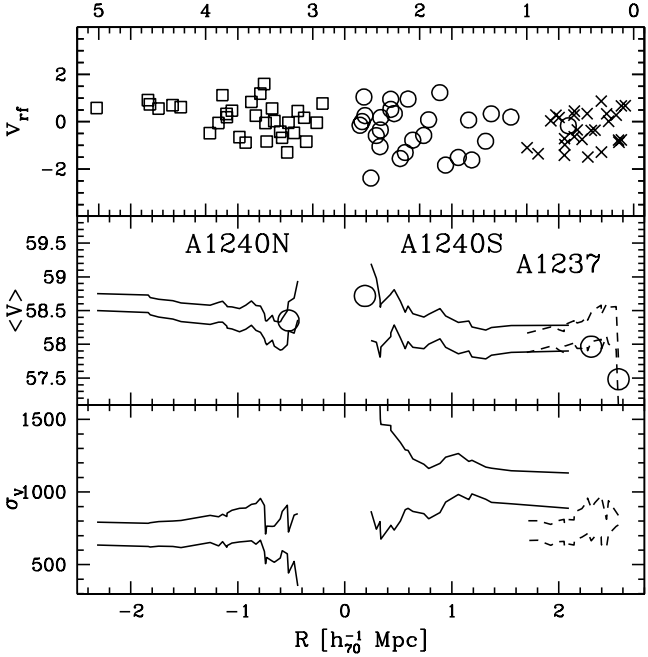}}
\caption
{ {\em Upper panel}: rest--frame velocity (in units of $10^3$ \kss)
  vs. projected clustercentric distance for the members of A1240N
  (squares), A1240S (circles) and A1237 (crosses). In this and other
  panels the lower axis give the clustercentric distance of A1240N
  (A1240S) galaxies versus the negative (positive) axis from the A1240
  center, while the upper axis give the clustercentric distance of
  A1237 galaxies from the A1237 center. The center of A1240 is defined
  as the X--ray center. The center of A1237 is defined as the position
  of the main peak in the 2D analysis. {\em Central and lower panels}:
  $68\%$ error bands of integral profiles of LOS mean velocity (in
  units of $10^3$ \kss) and LOS velocity dispersion (in units of \kss)
  for A1240N and A1240S (solid lines) and A1237 (dashed lines).  The
  circles in the {\em central panel} indicate the positions and
  velocities of the two brightest galaxies of A1240 (BCGN and BCGS)
  and the two brightest galaxies of A1237 (BCG1 and BCG2).}
\label{figprofnew}
\end{figure}

Figure~\ref{figprofnew} compares the $\left<\rm{v}\right>$ and
$\sigma_{\rm V}$ profiles of A1240 and A1237, where for A1240 we show
separately the results for A1240N and A1240S. The value 
$\left<\rm{v}\right>$ of A1237 is similar to that of A1240S,
suggesting a continuity in the velocity field.  However the
value of $\sigma_{\rm V}$ of A1237 is clearly lower than that of A1240S,
suggesting that A1237 is really a less massive system.

\subsection{Internal structure of A1240}
\label{int}

According to the 1DKS--test there is no difference between the
velocity distributions of A1240N and A1240S. The velocity distribution
of A1240 shows signatures of non--Gaussianity similar to those of the
whole cluster complex (e.g. a slight asymmetry and an important gap,
see Fig.~\ref{figstrip}). Like for the A1237 vs. A1240 case, we find
that A1240N and A1240S are so similar in the velocity space that the
velocity information is not useful to improve the galaxy
assignment. However, at the contrary of the A1237 vs. A1240 case,
A1240N and A1240S are spatially closer and likely strongly interacting
(see Sect.~\ref{disc}). This suggests that the galaxy assignment might
be questionable and that we must devote more cure in determining the
individual kinematical properties of the two subclumps.

\begin{figure}
\centering \resizebox{\hsize}{!}{\includegraphics{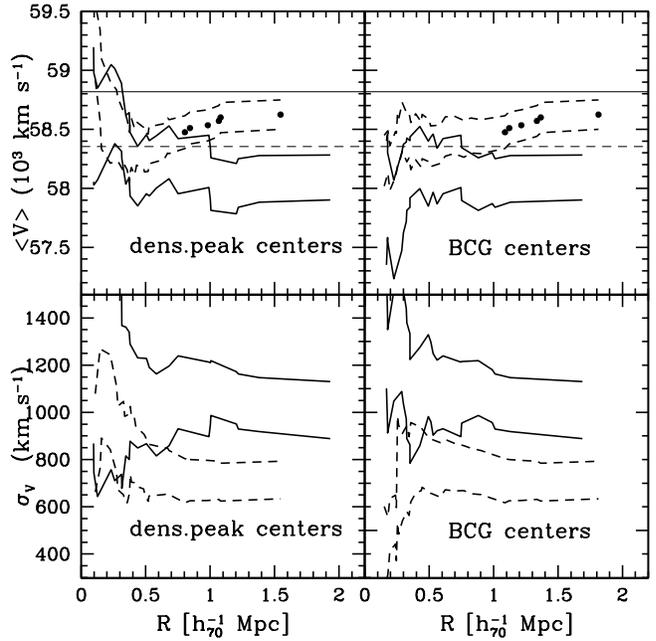}}
\caption
{ {\em Upper panels}: Error bands of integral profiles of LOS
  mean velocity for A1240N (dashed line) and A1240S (solid line). The
  distance is computed using as the center the position of the
  most important peak (on the left) and the position of the
  brightest cluster member (on the right). The horizontal lines
  indicate the velocities of BCGN and BCGS (dashed and solid lines,
  respectively). The close circles point out at which radius the six
  northern high velocity galaxies in A1240N are included to show
  their effect on the estimate of the mean velocity (see the text
  and Fig.~\ref{figvel}, too). {\em Lower panels}: The same for
  integral profiles of the LOS velocity dispersion.}
\label{figprofcori}
\end{figure}

The value of global $\left<\rm{v}\right>$ of A1240N is nominally
larger than that of A1240S, i.e.  $\left<\rm{v}\right>=(58624\pm 128)$
\ks and $\left<\rm{v}\right>=(58091\pm 195)$ \ks respectively, only at
a $\sim$ 2$\sigma$ c.l.  Moreover, looking at Fig.~\ref{figprofnew},
the two cores of A1240N and A1240S seem to have similar
$\left<\rm{v}\right>$: this is better shown in Fig.\ref{figprofcori},
where we directly compare the profiles of A1240N and A1240S.  The
large global $\left<\rm{v}\right>$ of A1240N is likely due to few high
velocity galaxies in the extreme northern cluster regions (see
the six galaxies shown as close circles in Fig.~\ref{figvel} -- right
panel; see also Fig.~\ref{figprofcori}). Indeed, the recent
merger of two subclumps may result in a plume, or arm, of outflying
galaxies detected for their different velocity with respect to the
cluster (see e.g. the case of Abell 3266; Quintana et al. \cite{qui96}
and Flores et al. \cite{flo00}).  Due to these difficulties in
detecting a quantitative difference between A1240N and A1240S, we
prefer to use the position and velocity of BCGN and BCGS as tracers of
the two interveining clumps. In fact, dominant galaxies trace the
cluster substructures (Beers \& Geller \cite{bee83}) and are good
tracers of interacting subclumps during a cluster merger, too (e.g.,
Boschin et al. \cite{bos06}; Barrena et al. \cite{bar07a}; Boschin et
al. \cite{bos09}).

The nominal value of global $\sigma_{\rm V}$ of A1240N is smaller than
that of A1240S, $\sigma_{\rm V}=709_{-83}^{+88}$ \ks and $\sigma_{\rm
  V}=991_{-99}^{+149}$ \ks respectively.  Using the brightest galaxies
as centers the comparison of the respective $\sigma_{\rm V}$ profiles
confirms this trend (see Fig.\ref{figprofcori} -- right lower panel).
Thus, although the nominal values of individual $\sigma_{\rm V}$ might
be not fully reliable, we decide to adopt them for describing A1240N
and A1240S.

Table~\ref{tabv} summarizes the properties of A1240N
and A1240S.

\subsection{Internal structure of A1237}
\label{int2}

As for A1237, it appears dominated by two bright galaxies, BCG1 and
BCG2 (IDs.~29 and 27, see Table~\ref{catalogA1240}) and the 2D DEDICA
algorithm shows the presence of one significant peak (at the
  $>99\%$ c.l.)  and two minor peaks significant at the $97\%$ and
  $89\%$ c.l. (see the somewhat elongated structure of A1237 in
  Fig.~\ref{figk2z}). Both the analysis of SDSS and our photometric
  data show that A1237 has only one well defined peak in the 2D galaxy
  distribution (see Sect.~\ref{2D}).  This last result is based on a
  much larger sample, thus we conclude that we have no evidence for a
  complex structure in A1237. The velocity distribution shows
evidence for a platykurtic behavior, but at a poorly significant
level (at the $\sim 95\%$ c.l.).

\subsection{Mass estimates}
\label{mass}

Under the usual assumptions (cluster sphericity, dynamical equilibrium
and galaxy distribution tracing the mass distribution), we can compute
global virial quantities.  Following the prescriptions of Girardi \&
Mezzetti (\cite{gir01}), the virial radius is $R_{\rm vir}=0.17\times
\sigma_{\rm V}/H(z)$ \h (see their Eq.~1 after introducing the scaling
with $H(z)$; see also Eq.~ 8 of Carlberg et al. \cite{car97} for
$R_{200}$) and the virial mass (Limber \& Mathews \cite{lim60}; see
also, e.g., Girardi et al. \cite{gir98}) is:

\begin{equation}
M=3\pi/2 \cdot \sigma_{\rm V}^2 R_{\rm PV}/G-{\rm SPT}.
\end{equation}

\noindent The quantity SPT, the surface pressure term correction (The
\& White \cite{the86}), is assumed to be equal to the $20\%$ of
the mass since this is the typical value recovered for the mass
computed within the virial radius in the literature when the data of
many clusters are combined together to enlarge the statistics
(Carlberg et al. \cite{car97}; Girardi et al. \cite{gir98}). The
quantity $R_{\rm PV}$ is a projected radius (equal to two times the
projected harmonic radius).  The value of $R_{\rm PV}$ depends on the
size of the sampled region and possibly on the quality of the spatial
sampling (e.g., whether the cluster is uniformly sampled or not).  It
is also possible to use an alternative estimate of $R_{\rm PV}$ based
on a priori knowledge of the galaxy distribution (see the
Appendix in Girardi et al. \cite{gir95}). Following Girardi et
al. (\cite{gir98}; see also Girardi \& Mezzetti \cite{gir01}) we can
assume a King--like distribution with parameters typical of
nearby/medium--redshift clusters: a core radius $R_{\rm c}=1/20\times
R_{\rm vir}$ and a slope--parameter $\beta_{\rm fit,gal}=0.8$, i.e.,
the volume galaxy density at large radii scales as $r^{-3 \beta_{\rm
fit,gal}}=r^{-2.4}$. With these assumptions we can use the
eq.~A6 of Girardi et al. (\cite{gir95}) to estimate $R_{\rm PV}$,
see also eq.~13 of Girardi et al. \cite{gir98} for a useful
approximation (i.e.,  $R_{\rm PV}=1.189 R_{\rm vir}[1+0.053(R_{\rm
vir}/R_{\rm c})]/ [1+0.117(R_{\rm vir}/R_{\rm c})]$.  Having
assumed the galaxy distribution, the value of $R_{\rm PV}$ depends 
only on $R_{\rm vir}$. In this way our estimates of global
virial quantities only depend on our estimate of $\sigma_{\rm V}$.

Through this procedure we obtain a mass estimate $M_{\rm A1237}(<R_{\rm
  vir}=1.63 \hhh)=(6\pm2)$ \mqua and $M_{\rm A1240}(<R_{\rm
  vir}=1.92 \hhh)=(9\pm3)$ \mquaa.

When a cluster shows a strongly substructured appearance (e.g., a
bimodal structure), the use of the global $\sigma_{\rm V}$ to compute
the mass might be misleading (Girardi et al. \cite{gir97} and
refs. therein). The true mass could be overestimated or
  underestimated depending on the angle of view of the cluster
  structure.  When the two subclumps are aligned along the LOS, they
  cannot be clearly distinguished in their projection onto the sky, but
  they can appear as two peaks (less or more overlapped, depending on
  their relative velocity) in the redshift distribution: in this case
  the mass estimated by the observed global $\sigma_{\rm V}$ is
  likely to be an overestimate of the true cluster mass (e.g.,
  cf. Tabs.~7 and ~8 of Girardi et al.  \cite{gir98}). When the two
  subclumps are aligned along an axis parallel to the plane of sky,
  they appear as two structures in their projection onto the sky but
  they cannot be distinguished in the redshift distribution (since
  their relative velocity has no component along the LOS
  direction). In this case the global velocity distribution is likely
  formed by the velocity distributions of the two clusters somewhat
  superimposed and the observed global $\sigma_{\rm V}$ does not take
  into account the existence of both the two subclumps.

The A1240N+A1240S system is comparable to the second case discussed
above: the two subclumps are distinguished in the sky, but lie at a
similar $z$. Indeed our analysis of Sect.~\ref{disc} will show that
the axis of the A1240N+A1240S system is likely roughly parallel to the
plane of sky.  A more reliable mass estimate of A1240 might be
obtained adding the mass estimates of the two subclumps, each mass
recovered from their respective $\sigma_{\rm V}$ (see
Table~\ref{tabv}).  We obtain a mass $M_{\rm A1240}\sim 1.9$ \mquii.
Another possible approach is to consider the future, virialized A1240
cluster and its global properties.  Since the cluster virial mass
computed within the virial radius scales with $\sim \sigma_{\rm V}^3$,
we expect that the $\sigma_{\rm V,vir}$ of the virialized A1240
corresponds to $(\sigma_{\rm V,A1240N}^3+\sigma_{\rm
V,A1240S}^3)^{1/3}$, i.e. $\sigma_{\rm V,vir} \sim 1100$ \ks well
larger than the LOS $\sigma_{\rm V}$ measured on observed data. This
corresponds to a mass of $M_{\rm A1240}(<R_{\rm vir}=2.4 \hhh)\sim
1.9$ \mqui in good agreement with the above estimate.

In conclusion, we estimate a mass range of $M_{\rm A1240}=(0.9-1.9)$
\mqui and of $M_{\rm A1237+A1240}=(1.5-2.5)$ \mquii.

\subsection{Analysis of photometric data}
\label{2D}

The results of the 2D DEDICA method applied to the 89 cluster members
are shown in Sect.~\ref{memb}. However, our spectroscopic data suffer
from magnitude incompleteness and the field around the cluster is not
covered in an homogeneous way. To overcome these limits we use our
photometric catalog.

\begin{figure}
\centering
\includegraphics[width=8cm]{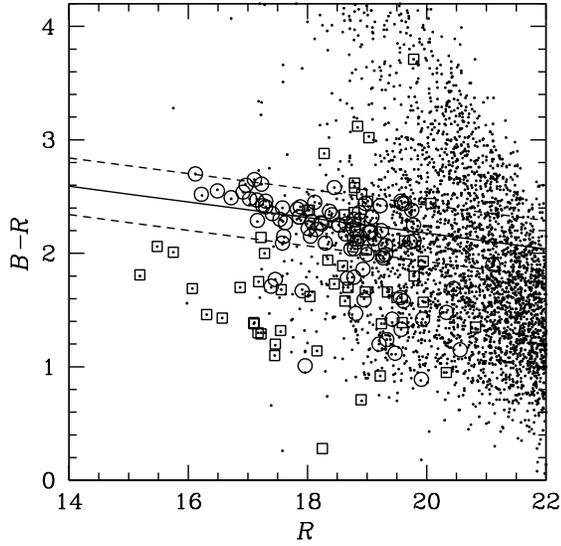}
\caption
{$B$--$R$ vs. $R$ diagram for galaxies with available spectroscopy is
  shown by large circles and squares (members and non members,
  respectively).  Small points indicate galaxies found in our INT
  photometric sample, i.e. objects with a class star $\leq 0.9$.  The
  solid line gives the best--fit color--magnitude relation as
  determined on spectroscopically confirmed member galaxies; the
  dashed lines are drawn at $\pm$0.25 mag from the color--magnitude
  relation.}
\label{figcm}
\end{figure}

We select likely members on the basis of the $B$--$R$ vs. $R$ relation,
as already performed in some previous works of ours (e.g., Barrena et
al. \cite{bar07b}). To determine the relation we fix the slope
according to L\'opez--Cruz et al. (\cite{lop04}, see their Fig.~3) and
apply the two--sigma--clipping fitting procedure to the cluster
members obtaining $B$--$R=3.557-0.069\times R$ for the red sequence of
A1240 galaxies. Figure~\ref{figcm} shows $B$--$R$ vs. $R$ diagram for
galaxies with available spectroscopy and the fitted line. 

\begin{figure}
\includegraphics[width=8cm]{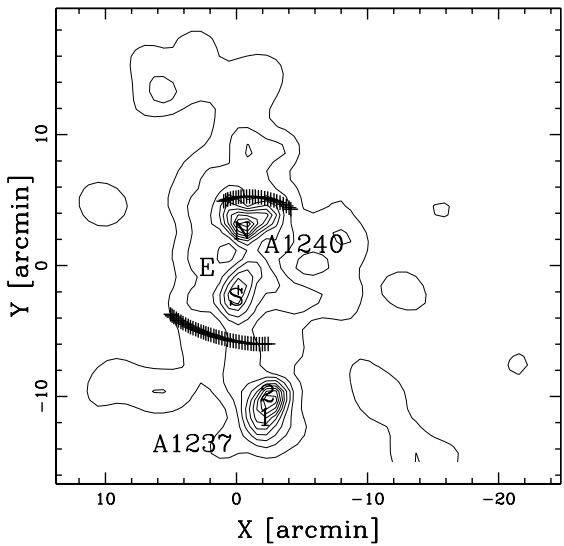}
\caption
{Spatial distribution on the sky and relative isodensity contour map
  of likely cluster members (applying the color--magnitude selection
  to the INT photometric catalog) with $R\le 20.5$, obtained with the
  DEDICA method.  The positions of the brightest galaxies are
  indicated. The two relics  are indicated in a schematic way.}
\label{figk2cm}
\end{figure}

Out of our photometric catalog we consider galaxies (objects with
SExtractor stellar index $\le 0.9$) lying within 0.25 mag of the
relation. Following Visvanathan \& Sandage (\cite{vis77}) the width
  of 0.25 mag approximately corresponds to 2$\sigma$ around the
  color--magnitude relation (see also Mazure et al. \cite{maz88} for
  a classical application to Coma cluster).  The contour map for 370 likely
cluster members having $R\le 20.5$ shows the bimodal structure of
A1240 and the presence of A1237, confirming the results of
Sect.~\ref{memb} (see Fig.~\ref{figk2cm}). Similar results are
obtained with different magnitude cut--offs (e.g., $R\le 20$ and $R\le
21$).

In this paper we also use public photometric data from the SDSS. In
particular, we use $r'$, $i'$, $z'$ magnitudes, already corrected for
Galactic extinction and consider galaxies within a radius of 30\arcm
from the cluster center.

\begin{figure}
\includegraphics[width=8cm]{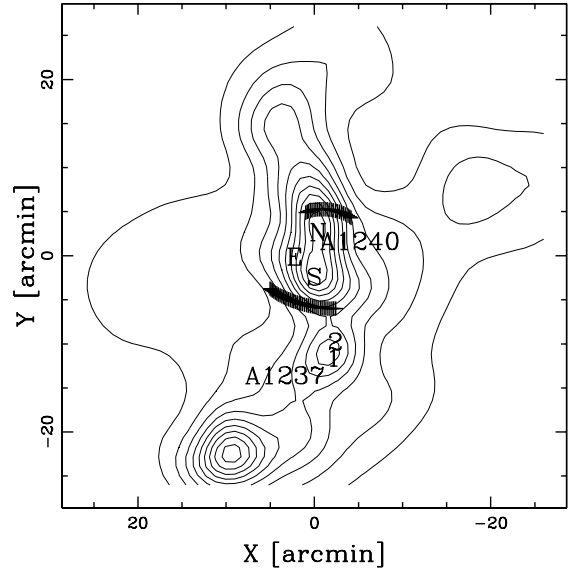}
\caption
{Spatial distribution on the sky and relative isodensity contour map
  of likely cluster members (applying the color--color selection to
  the SDSS photometric catalog) with $r'\le 20.8$, obtained with the
  DEDICA method.  The positions of the brightest galaxies are
  indicated. The two relics  are indicated in a schematic way.}
\label{figk2metacc}
\end{figure}

Following Boschin et al. (\cite{bos08}, see also Goto et
al. \cite{got02}), out of SDSS photometric catalog we consider
galaxies (here objects with $r'$ phototype parameter $=3$) lying
within $\pm$0.08 mag from the median values of $r'$--$i'$=0.47 and
$i'$--$z'$=0.32 colors of the spectroscopically cluster members. The
value of 0.08 mag is two times the typical scatter reported by Goto et
al. (\cite{got02}) for the corresponding color--magnitude relations
$r'$--$i'$ vs. $r'$ and $i'$--$z'$ vs. $r'$.  However, this
member selection seems not enough conservative for the case of
A1240. In fact, using our spectroscopic catalog, we notice that this
color--color selection recognizes as ``likely members'' 21 out of the
56 non member galaxies (against the 11 out of 56 by using the
color--magnitude selection). Therefore we decide to use here a more
conservative selection, i.e. considering only galaxies lying within
$\pm$0.04 mag from the median values of $r'$--$i'$ and $i'$--$z'$
colors (see Fig.~\ref{figk2metacc} with 272 galaxies with $r'<20.8$).
The galaxy distribution shows the N--S elongation of A1240 and the
presence of A1237. Moreover, another cluster is shown at $\sim$
25\arcm South--East of A1240. This cluster was also detected by Gal et
al. (\cite{gal03}) and Koester et al. (\cite{koe07}), who estimated a
photometric $z \sim 0.18$--0.19.

\section{X--ray analysis of A1240}
\label{xray}

The X--ray analysis of A1240 is performed on the archival data of the
Chandra ACIS--I observation 800407 (exposure ID \#4961). The pointing
has an exposure time of 52 ks. Data reduction is performed using the
package CIAO\footnote{CIAO is freely available at
http://asc.harvard.edu/ciao/} (Chandra Interactive Analysis of
Observations) on chips I0, I1, I2 and I3 (field of view $\sim
17\arcmin\times 17\arcmin$). First, we remove events from the level 2
event list with a status not equal to zero and with grades one, five
and seven. Then, we select all events with energy between 0.3 and 10
keV. In addition, we clean bad offsets and examine the data, filtering
out bad columns and removing times when the count rate exceeds three
standard deviations from the mean count rate per 3.3 s interval. We
then clean the chips for flickering pixels, i.e., times where a pixel
has events in two sequential 3.3 s intervals. The resulting exposure
time for the reduced data is 51.3 ks.

A quick look of the reduced image is sufficient to have an hint of the
morphology of the extended X--ray emission of A1240. However, the
cluster was centered exactly in the ``cross'' of the gaps among the
four chips. To obtain a more precise result, we have to correct the
photon counts in the poorly exposed ACIS CCD gaps. First, we operate a
binning of the reduced image. Then, we operate a soft smoothing and
correct the photon counts with a division by an exposure map. The
result is an image from which we extract the contour levels (soft
photons in the energy range 0.5--2 keV) plotted in
Fig.~\ref{figimage}.

The complex X--ray morphology of A1240 is evident. In particular, the
central X--ray emission is clearly elongated in the N--S direction,
i.e. the same direction defined by the two galaxy clumps A1240N and
A1240S. Contour levels in Fig.~\ref{figimage} also reveal a diffuse
source $\sim$6\arcm SE of the cluster center. A visual inspection of
the INT $R$--band image shows that this structure does not match with
any evident galaxy concentration. We suspect it could be a distant
background galaxy group. Unfortunately, a more quantitative
morphological analysis is not trivial. In fact, the gaps among the
ACIS chips make quite critical the computation of the surface
brightness distribution. Moreover, the disturbed morphology of the ICM
does not justify the spherical symmetric assumption. This does not
encourage the presentation of X-ray profiles.

\begin{figure}
\centering
\resizebox{\hsize}{!}{\includegraphics{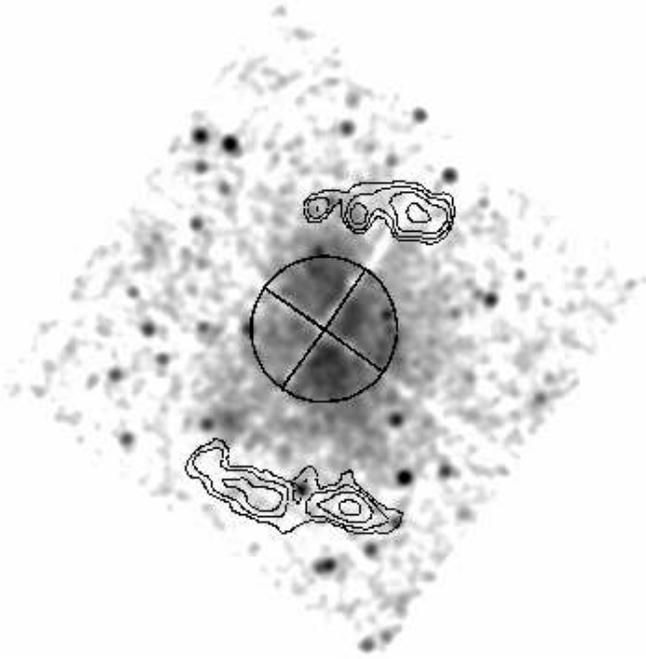}}
\caption
{17\arcmm$\times$17\arcm Chandra X--ray {\bf smoothed} image (ID~4961)
of A1240 in the energy band 0.5--2 keV (North at the top and East to
the left). The central circle has a radius of $\sim$2.58\arcm (0.5 Mpc
at the cluster redshift, see text). Black contours show the two radio
relics.}
\label{figX}
\end{figure}

As for the spectral properties of the cluster X--ray photons, we
compute a global estimate of the ICM temperature. The temperature is
computed from the X--ray spectrum of the cluster within a circular
aperture of $\sim$2.58$\arcmin$ radius (0.5 Mpc at the cluster
redshift; see Fig.~\ref{figX}) around the center of the four
ACIS chips. Fixing the absorbing galactic hydrogen column density at
1.99$\times$10$^{20}$ cm$^{-2}$, computed from the HI maps by Dickey
\& Lockman (\cite{dic90}), we fit a Raymond--Smith (\cite{ray77})
spectrum using the CIAO package Sherpa with a $\chi^{2}$ statistics
and assuming a metal abundance of 0.3 in solar units. We find a best
fitting temperature of $T_{\rm X}=\,$6.0\,$\pm\,0.5$ keV.

We then search for temperature gradients by dividing the 0.5 Mpc
radius circle in four quadrants (chosen in order to avoid the
gaps among the chips), as shown in Fig.~\ref{figX}. As expected for a
merging cluster, there is some evidence that the ICM temperature is
not homogeneous, with the Western quadrant ($T_{\rm
X}$=6.7$^{+1.9}_{-1.2}$ keV) somewhat hotter than the Northern
($T_{\rm X}$=5.1$^{+1.0}_{-0.8}$ keV), the Eastern ($T_{\rm
X}$=5.2$^{+1.9}_{-1.2}$ keV) and Southern ($T_{\rm
X}$=5.4$^{+0.9}_{-0.8}$ keV) quadrants. More detailed temperature
and metallicity maps would be highly desirable to better describe the
properties of the ICM, but the photon statistics is not good enough to
this aim. In particular, we do not provide any estimate of the ICM
temperature in proximity of the two relics, where the X--ray surface
brightness is too low to obtain any reliable measurement (see
Fig.~\ref{figX}).

\section{Cluster structure: discussion}
\label{disc}

Both optical and X--ray data indicate that A1240 is a strongly
substructured cluster, elongated in the N--S direction, the same of
the axis of symmetry of the relics.  We also detect two clumps,
separated by $1.2$ \hh, in the galaxy distribution. These
observational features suggest that we are looking at a cluster just
forming through the merging of two main subclumps. The difficulty in
separating the two subclumps in the velocity space and the small LOS
velocity difference of the two BCGs suggest that the axis of the
merger lies in the plane of the sky. The evidence of a very disturbed
ICM distribution, somewhat displaced from the galaxy
distribution, suggests that the merger is seen after the phase of the
core passage, as shown by the results of the numerical simulation
(e.g., Roettiger et al. \cite{roe97}, their Figs. 7--14).

As for the observational cluster structure, A1240 is similar to Abell
3667 (see Roettiger et al. \cite{roe99} and refs. therein) where the
optical and X--ray structures are elongated in a direction roughly
similar to that of the axis of symmetry of the two radio relics which
are separated by $\sim 3-4$ \hh. The two intervening galaxy subclumps
are separated by $\sim 1$ \h with a small LOS velocity difference
$\sim 120$ \ks between the dominant galaxies.  Basic observational
features of Abell 3667 were explained with the ``outgoing merger
shocks'' model, where shocks provide sites for diffusive shock
acceleration of relativistic electrons causing the presence of the
radio sources (Roettiger et al. \cite{roe99}).

For A1240 we have detected two galaxy subclumps in the N--S direction,
the same direction of the elongation of the X--ray surface brightness
and the axis of symmetry of the two radio relics. The values of
relevant parameters for the two--clumps system, as deduced from the
BCGN and BCGS, are: the relative LOS velocity in the rest--frame,
$V_{\rm rf}=390$ \ks, and the projected linear distance between the
two clumps, $D=1.2$ \hh. The two roughly symmetric relics lie more
externally, separated by a projected linear distance $\sim 2$ \h
(Bonafede et al. \cite{bon09}).

We now use the above parameters and the mass of the system computed in
a range $M_{\rm A1240}=(0.9-1.9)$ \mqui to investigate the relative
dynamics of A1240N and A1240S. In particular, we use different
analytic approaches based on an energy integral formalism in the
framework of locally flat spacetime and Newtonian gravity (e.g., Beers
et al. \cite{bee82}).

\begin{figure}
\centering
\resizebox{\hsize}{!}{\includegraphics{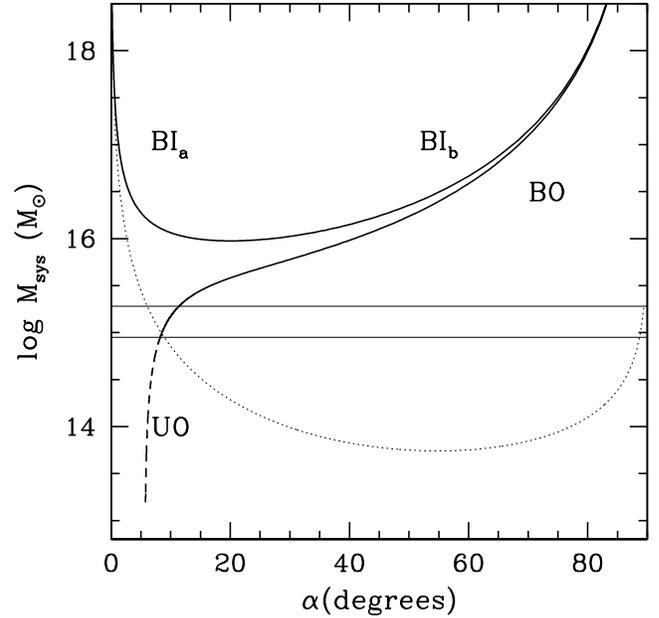}}
\caption
{System mass vs. projection angle for bound and unbound solutions
  (thick solid and thick dashed curves, respectively) of
  the two--body model applied to A1240N and A1240S subsystems.  Labels
  BI$_{\rm a}$ and BI$_{\rm b}$ indicate the bound and incoming,
  i.e. collapsing solutions (thick solid curve).  Label BO
  indicates the bound outgoing, i.e. expanding solutions (thick
  solid curve).  Label UO indicates the unbound outgoing solutions
  (thick dashed curve).  The horizontal lines give the range of
  observational values of the mass system.  The thin dashed
  curve separates bound and unbound regions according to the Newtonian
  criterion (above and below the thin dashed curve,
  respectively).  }
\label{figbim}
\end{figure}

First, we compute the Newtonian criterion for gravitational binding
stated in terms of the observables as $V_{\rm r}^2D\leq2GM_{\rm
sys}\rm{sin}^2\alpha\,\rm{cos}\alpha$, where $\alpha$ is the
projection angle between the plane of the sky and the line connecting
the centers of the two clumps. The thin curve in Fig.~\ref{figbim}
separates the bound and unbound regions according to the Newtonian
criterion (above and below the curve, respectively). Considering the
lower (upper) limit of $M_{\rm A1240}$, the N+S system is bound between
$9\degree$ and $84\degree$ ($6\degree$ and $89\degree$); the
corresponding probability, computed considering the solid angles
(i.e., $\int^{\alpha2}_{\alpha1} {\rm cos}\,\alpha\,d\alpha$), is 84\%
(89\%).

Then, we apply the analytical two--body model introduced by Beers et
al. (\cite{bee82}) and Thompson (\cite{tho82}; see also Lubin et
al. \cite{lub98} for a more recent application). This model assumes
radial orbits for the clumps with no shear or net rotation of the
system. According to the boundary conditions usually considered, the
clumps are assumed to start their evolution at time $t_0=0$ with
separation $d_0=0$, and now moving apart or coming together for the
first time in their history.

In the case of A1240N+A1240S system, where the first core passage has
likely already occurred, we assume that the time $t_0=0$ with
separation $d_0=0$ is the time of their core crossing and that we are
looking at the cluster a time $t$ after.  To obtain an estimate of $t$
we use the Mach number of the shock ${\cal M}\sim 3$ as recovered by
Bonafede et al. (\cite{bon09}) from the radio spectral index.  The
Mach number is defined as ${\cal M}={\rm v}_{\rm s}/c_{\rm s}$, where
${\rm v}_{\rm s}$ is the velocity of the shock and $c_{\rm s}$ in the
sound speed in the pre--shock gas (see e.g., Sarazin \cite{sar02} for
a review). The value of $c_{\rm s}$, obtained from our estimate of
$T_{\rm X}$, i.e. $c_{\rm s}\sim 995$ \kss, leads to a value of ${\rm
v}_{\rm s}\sim 3\times 10^3$ \kss. Assuming the shock velocity as a
constant, the shock covered a $\sim 1$ \h scale (i.e. the distance of
the relics from the cluster center) in a time of $\sim 0.3$ Gyrs. We
assume this time as our estimate of $t$. Although the velocity of the
shock is not constant, recent studies based on numerical simulations
show how the variation of ${\rm v}_{\rm s}$ is much smaller than the
variation of the relative velocity of the subclumps identified with
their dark matter components (see Fig.~4 of Springel \& Farrar
\cite{spr07}; see Fig.~14 of Mastropietro \& Burkert \cite{mas08}),
thus our rough estimate of $t$ is acceptable at the first order of
approximation.

The bimodal model solution gives the total system mass $M_{\rm sys}$
as a function of $\alpha$ (e.g., Gregory \& Thompson \cite{gre84}).
Figure~\ref{figbim} compares the bimodal--model solutions with the
observed mass of the system.  The present solutions span the bound
outgoing solutions (i.e. expanding), BO; the bound incoming solutions
(i.e. collapsing), BI$_{\rm a}$ and BI$_{\rm b}$; and the unbound
outgoing solutions, UO.  For the incoming case there are two solutions
because of the ambiguity in the projection angle $\alpha$. Both the BO
and UO solutions are, in principle, consistent with the observed mass
range. However, the BO solution is the more likely since the
probability associated to the BO solution is much higher than that
associated to the UO solution. In fact, we obtain that P(BO) and
P(UO) are $92\%$ and $8\%$, respectively, where these probabilities
are computed considering the solid angles (see above) and assuming
that the region of $M_{\rm sys}$ values is equally probable for
individual solutions (see e.g.  Barrena et al. \cite{bar07b}).  As
for the projection angle we estimate a value of $\alpha\sim
10$\degree.  This small angle means that the real spatial distance
between A1240 subclumps is similar to the projected one while the
real, i.e. deprojected, velocity difference is $V_{\rm rf} \sim 2000$
\kss, a quite reasonable value during cluster mergers (see,
e.g. Springel \& Farrar \cite{spr07} and refs. therein).  Notice that
the relative velocity between galaxy clumps is smaller than the shock
velocity, i.e.  the regime is not stationary, but this is expected
when comparing shock and collisionless components in numerical
simulations (Springel \& Farrar \cite{spr07}; Mastropietro \& Burkert
\cite{mas08}).

Since the merger of A1240 clumps occurred largely in the plane of the
sky, this likely explains the similarity between the observational
features of A1240 and Abell 3667. However, the case of A1240 is made
more complex by the presence of A1237.

We investigate the relative dynamics of A1237 and A1240 with the same
approach described above. The values of the relevant observable
quantities for the two--clumps system are based on Table~\ref{tabv},
i.e. LOS $V_{\rm rf}=320$ \ks and $D=2.7$ \hh. For the mass of the
system we use the mass range computed in Sect.~\ref{mass}, $M_{\rm
A1237+A1240}=(15-25)$ \mquii. The Newtonian criterion for
gravitational binding leads to a system bound between $8\degree$ and
$89\degree$ ($7\degree$ and $89\degree$); the corresponding
probability, computed considering the solid angles (i.e.,
$\int^{89}_{6} {\rm cos}\,\alpha\,d\alpha$), is $87\%$ ($89\%$).

\begin{figure}
\centering
\resizebox{\hsize}{!}{\includegraphics{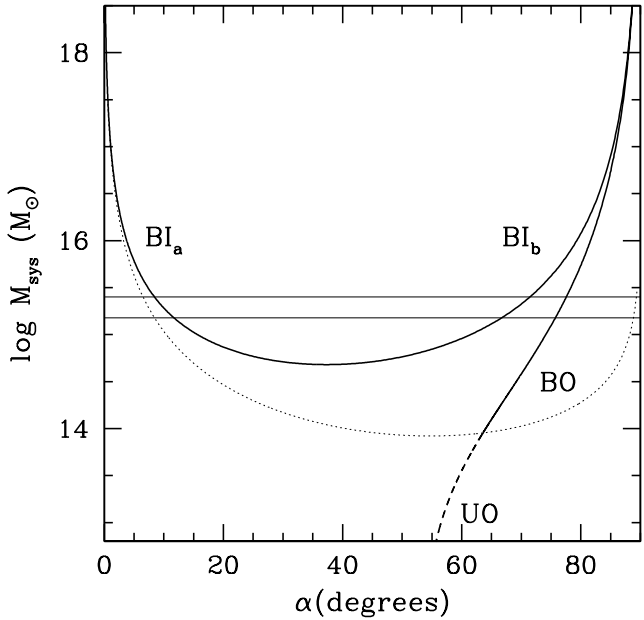}}
\caption
{System mass vs. projection angle for bound and unbound solutions
  (solid and dashed curves, respectively) of the two--body model
  applied to A1237 and A1240 systems.}  
\label{figbimss}
\end{figure}

  In the case of A1237 and A1240 we have no evidence of a previous
  merger since the gas distribution of A1240 (although very complex)
  does not show evidence of a peculiar displacement in direction of
  A1237.  Therefore we assume that we are looking at A1237 and A1240
  before their encounter.  X--ray observations of A1237 would be
  useful to confirm this hypothesis.  Under our assumption we can use
  the standard version of the analytical two--body model to study the
  A1237+A1240 system, where the clumps are assumed to start their
  evolution at time $t_0=0$ with separation $d_0=0$, and are moving
  apart or coming together for the first time in their history.  We
  are looking at the system at the time of $t=11.090$ Gyrs, i.e. the
  age of the Universe at the system redshift.  The solutions
  consistent with the observed mass span these cases: the bound and
  present incoming solution (i.e. collapsing), BI$_{\rm a}$ and
  BI$_{\rm b}$, and the bound--outgoing solution, BO (see
  Fig.~\ref{figbimss}). The associated probabilities are $61\%$,
  $31\%$ and $8\%$, for BI$_{\rm a}$, BI$_{\rm b}$ and BO
  respectively.  The most probable solution is also the most
  interesting one since the BI$_{\rm a}$ solution gives the same
  value of $\alpha\sim 10$\degree already found for the A1240N+S
  merger. In this case, the direction of the clump velocities
  are consistent, too: at the present time A1240N has already crossed
  A1240S, being now in front of A1240S and moving outgoing; A1237 is
  lying behind A1240 and is infalling onto it along almost the same
  N--S direction.

\section{Conclusions}
\label{conc}

Our results strongly support the conclusion of Bonafede et
al. (\cite{bon09}), based on radio data, in favor of the ``outgoing
merger shocks'' model for the double relics of A1240. In fact, we
detect the intervening merging subclumps and recover an acceptable
model for the internal cluster dynamics.  Moreover, our results also
suggest that we are looking at the cluster accretion along a large
scale structure filament.

Our analysis shows how powerful is the study of the internal cluster
dynamics through the analysis of kinematics of cluster galaxies.
Other insights into A1240 might be recovered from a better knowledge
of galaxy properties, e.g.  spectral signatures of past activity which
could be useful to determine the relevant time--scales (see e.g.,
Ferrari et al. \cite{fer03}; Boschin et al. \cite{bos04}) and from
deeper X--ray observations (see, e.g Vikhlinin et al. \cite{vik01} for
the discover of a ``cold front'' in Abell 3667).  Finally, we point
out that A1240, with its displacement between collisional (gas) and
collisionless components (galaxies), is an excellent candidate to
study the properties of the dark matter, in particular its collisional
cross section, as already performed in other merging clusters
(Markevitch et al. \cite{mar04}; Brad\v{a}c et al. \cite{bra08}).

\begin{acknowledgements}
We are in debt with Klaus Dolag for interesting discussions.  We thank
the anonymous referee for her/his useful suggestions. This publication
is based on observations made on the island of La Palma with the
Italian Telescopio Nazionale Galileo (TNG) and the Isaac Newton
Telescope (INT). The TNG is operated by the Fundaci\'on Galileo
Galilei -- INAF (Istituto Nazionale di Astrofisica). The INT is
operated by the Isaac Newton Group. Both telescopes are located in the
Spanish Observatorio of the Roque de Los Muchachos of the Instituto de
Astrofisica de Canarias.

This research has made use of the NASA/IPAC Extragalactic Database
(NED), which is operated by the Jet Propulsion Laboratory, California
Institute of Technology, under contract with the National Aeronautics
and Space Administration.

This research has made use of the galaxy catalog of the Sloan Digital
Sky Survey (SDSS). Funding for the SDSS has been provided by the
Alfred P. Sloan Foundation, the Participating Institutions, the
National Aeronautics and Space Administration, the National Science
Foundation, the U.S. Department of Energy, the Japanese
Monbukagakusho, and the Max Planck Society. The SDSS Web site is
http://www.sdss.org/.

The SDSS is managed by the Astrophysical Research Consortium for the
Participating Institutions. The Participating Institutions are the
American Museum of Natural History, Astrophysical Institute Potsdam,
University of Basel, University of Cambridge, Case Western Reserve
University, University of Chicago, Drexel University, Fermilab, the
Institute for Advanced Study, the Japan Participation Group, Johns
Hopkins University, the Joint Institute for Nuclear Astrophysics, the
Kavli Institute for Particle Astrophysics and Cosmology, the Korean
Scientist Group, the Chinese Academy of Sciences (LAMOST), Los Alamos
National Laboratory, the Max--Planck--Institute for Astronomy (MPIA),
the Max--Planck--Institute for Astrophysics (MPA), New Mexico State
University, Ohio State University, University of Pittsburgh,
University of Portsmouth, Princeton University, the United States
Naval Observatory, and the University of Washington.

\end{acknowledgements}

\end{document}